\documentclass{vldb}
\usepackage{algorithmic}
\usepackage{algorithm}
\usepackage{subfigure}
\usepackage{epsfig}
\usepackage{multirow}
\usepackage{amsfonts}
\usepackage{array}
\usepackage{wrapfig}

\begin{document}

\title{Bayesian Locality Sensitive Hashing for Fast Similarity
Search}

\numberofauthors{1}
\author{
\alignauthor
Venu Satuluri
and
Srinivasan Parthasarathy\\
\affaddr{Dept. of Computer Science and Engineering}\\
\affaddr{The Ohio State University}\\
\email{\{satuluri,srini\}@cse.ohio-state.edu}
}

\maketitle

\begin{abstract}
Given a collection of objects and an associated similarity
measure, the all-pairs similarity search problem asks us to find
all pairs of objects with similarity greater than a certain
user-specified threshold. Locality-sensitive hashing (LSH) based methods 
have
become a very popular approach for this problem. However, most
such methods only use LSH for the first phase of similarity
search - i.e. efficient indexing for candidate
generation. In this paper, we present {\bf BayesLSH}, a principled Bayesian
algorithm for the subsequent phase of similarity search - performing candidate pruning and similarity
estimation using LSH. A simpler variant, BayesLSH-Lite, which
calculates similarities exactly, is also presented. BayesLSH is able to quickly prune away a large
majority of the false positive candidate pairs, leading to 
significant speedups over baseline
approaches. 
For BayesLSH,
we also provide probabilistic
guarantees on the quality of the output, both in terms of
accuracy and recall. Finally, the quality of BayesLSH's output
can be easily tuned and does not require any manual setting of
the number of hashes to use for similarity estimation, unlike
standard approaches. For two state-of-the-art candidate
generation algorithms, AllPairs~\cite{bayardo07} and LSH,
BayesLSH enables significant speedups, typically in the range
2x-20x for a wide variety of datasets.  
\end{abstract}

\section{Introduction}
\label{sec:intro}
Similarity search is a problem of fundamental importance for a
broad array of fields, including databases, data mining and
machine learning. The general problem is as follows: given a
collection of objects $D$ with some similarity measure $s$ defined between
them and a query object $q$, retrieve all objects from $D$ that
are most similar to $q$ according to the similarity measure $s$.
The user may be either interested in the top-$k$ most similar objects
to $q$, or the user may want all objects $x$ such that $s(x,q) >
t$, where $t$ is the similarity threshold. A more specific
version of similarity search is the \emph{All Pairs
similarity search} problem, where there is no explicit query
object, but instead the user is interested in all pairs of
objects with similarity greater than some threshold.
The number
of applications even for the more specific all pairs similarity
search problem is impressive:
clustering~\cite{ravichandran05}, semi-supervised
learning~\cite{zhu09}, information retrieval (including text, audio and
video), query refinement~\cite{bayardo07}, near-duplicate
detection~\cite{xiao08}, collaborative
filtering, link prediction for graphs~\cite{liben-nowell07}, and 3-D
scene reconstruction~\cite{agarwal09} among others. In many of
these applications, approximate solutions with 
small errors in similarity assessments are
acceptable if they can buy significant reductions in running time
e.g. in web-scale
clustering~\cite{broder97,ravichandran05}, information
retrieval~\cite{elsayed11},
near-duplicate detection for web
crawling~\cite{manku07,henzinger06} and
graph clustering~\cite{satuluri11}.

Roughly speaking, similarity search algorithms can be divided into two main phases
- \emph{candidate generation} and \emph{candidate verification}.
During the candidate generation phase, pairs of objects that are
good candidates for having similarity above the user-specified
threshold are generated using one or another indexing mechanism,
while during candidate verification, the similarity of each candidate
pair is verified against the threshold, in many cases by exact
computation of the similarity.
The traditional indexing structures used for candidate generation
were space-partitioning approaches such as kd-trees and R-trees,
but these approaches work well only in low dimensions (less
than 20 or so~\cite{datar04}). An important breakthrough 
was the invention of locality-sensitive
hashing~\cite{indyk98,gionis99},
where the idea is to find a family of hash functions such that
for a random hash function from this family, two objects with 
high similarity are very likely to be hashed to
the same bucket. One can then generate candidate pairs by hashing
each object several times using randomly chosen hash functions,
and generating all pairs of objects which have been hashed to the
same bucket by at least one hash function.
 Although LSH is a randomized, approximate
solution to candidate generation, 
similarity search based on LSH has nonetheless become
immensely popular because it provides a practical solution for
high dimensional applications along with theoretical guarantees
for the quality of the approximation~\cite{andoni08}.

In this article, we show how LSH can be exploited for the
phase of similarity search subsequent to candidate generation
i.e. candidate verification and similarity computation. We adopt
a principled Bayesian approach that allows us to reason about the
probability that a particular pair of objects will meet the user-specified
threshold by inspecting only a few hashes of each object, 
which in turn allows us to quickly prune away unpromising pairs.
Our Bayesian approach also allows us to estimate similarities to
a user-specified level of accuracy without requiring any tuning
of the number of hashes, overcoming a significant drawback
of standard similarity estimation using LSH. We develop two
algorithms, called {\bf BayesLSH} and {\bf BayesLSH-Lite}, where
the former performs both candidate pruning and similarity
estimation, while the latter only performs candidate pruning and
computes the similarities of unpruned candidates exactly. 
Essentially, 
BayesLSH provides a way to trade-off accuracy for
speed in a controlled manner. 
Both BayesLSH and BayesLSH-Lite can be 
combined with any existing candidate
generation algorithm, such as AllPairs~\cite{bayardo07} or LSH. 
Concretely, BayesLSH provides the following probabilistic
guarantees:

\smallskip 
\emph{\noindent
Given a collection of objects $D$, an associated similarity
function $s(.,.)$, and a similarity threshold $t$; recall
parameter $\epsilon$ and accuracy parameters $\delta, \gamma$; 
return pairs of objects $(x,y)$ along with 
similarity estimates $\hat{s}_{x,y}$ such that:
\begin{enumerate}
\item $Pr[ s(x,y) \geq t ] > \epsilon$ i.e. each pair with a
greater than $\epsilon$ probability of being a true positive is
included in the output set.
\item
$Pr[ |\hat{s}_{x,y} - s(x,y)| \geq \delta] < \gamma$ i.e. each
associated similarity estimate is accurate up to $\delta$-error
with probability $> 1-\gamma$.
\end{enumerate}
}

With BayesLSH-Lite, the similarity calculations are exact, so
there is no need for guarantee 2, but guarantee 1 from above stays. 
We note that the
parameterization of BayesLSH is intuitive - the desired recall
can be controlled using $\epsilon$, while $\delta, \gamma$
together specify the desired level of accuracy of similarity
estimation.

The advantages of BayesLSH are as follows:
\begin{enumerate}
\item
\vspace{-0.06in}
The general form of the algorithm can be easily adapted to work
for any similarity measure with an associated LSH family (see
Section~\ref{sec:bg} for a formal definition of LSH). We
demonstrate BayesLSH for Cosine and Jaccard similarity measures,
and believe that it can be adapted to other measures with LSH
families, such as kernel similarities.
\item
\vspace{-0.06in}
There are no restricting assumptions about the specific form of
the candidate generation algorithm; BayesLSH \emph{complements}
progress in candidate generation algorithms.
\item
\vspace{-0.06in}
For applications which already use LSH for candidate generation,
it is a natural fit since it exploits the hashes of the
objects for candidate pruning, further amortizing the costs of
hashing.
\item
\vspace{-0.06in}
It works for both binary and general real-valued vectors. This is
a significant advantage because recent progress in
similarity search has been limited to binary
vectors~\cite{xiao08,zhai11}. 
\item
\vspace{-0.06in}
Parameter tuning is easy and intuitive; the only parameters are
$\gamma, \delta$ and $\epsilon$, each of which, as we have seen,
directly control the quality of the output result. In particular,
there is no need for manually tuning the number of hashes, as one
needs to with standard similarity estimation using LSH. 
\end{enumerate}

We perform an extensive evaluation of our algorithms and comparison
with state-of-the-art methods, on a diverse array of 6 real datasets. We
combine BayesLSH and BayesLSH-Lite with two different candidate generation
algorithms AllPairs~\cite{bayardo07} and LSH, and find
significant speedups, typically in the range 2x-20x over baseline
approaches (see Table~\ref{tab:bestSpeedups}).
BayesLSH is able to achieve the speedups primarily by
being extremely effective at pruning away false positive
candidate pairs. To take a typical example,
BayesLSH is able to prune away 80\% of the
input candidate pairs after examining only 8 bytes worth of hashes per candidate
pair, and 99.98\% of the candidate pairs after examining
only 32 bytes per pair. Notably, BayesLSH is able to do such effective
pruning without adversely affecting the recall, which is still quite high,
generally at 97\% or above. Furthermore, the accuracy of
BayesLSH's similarity estimates is
much more consistent as compared to the standard similarity
approximation using LSH, which tends to produce
very error-ridden estimates for low similarities. Finally, we
find that parameter tuning for BayesLSH is intuitive and works as
expected, with higher accuracies and recalls being achieved
without leading to undue slow-downs.

\section{Background}
\label{sec:bg}
Following Charikar~\cite{charikar02}, we define a
locality-sensitive hashing scheme as a distribution on a family
of hash functions ${\cal F}$ operating on a collection of
objects, such that for any two objects ${\bf x}, {\bf y}$,
\begin{equation}
\label{eqn:lshmain}
Pr_{h \in {\cal F}} [h({\bf x}) = h({\bf y})] = sim({\bf x},{\bf y}) 
\end{equation}
It is important to note that the probability in
Eqn~\ref{eqn:lshmain} is for a random selection of the hash
function from the family ${\cal F}$. Specifically, it is
not for a random pair ${\bf x},{\bf y}$ - i.e. the equation is valid for
\emph{any} pair of objects ${\bf x}$ and ${\bf y}$. The output of the hash
functions may be either bits (0 or 1), or integers.
Note that this definition of LSH, taken from~\cite{charikar02},
is geared towards similarity 
measures and is more useful in our context, as compared to the
slightly different definition of LSH used by many other
sources~\cite{datar04,andoni08}, including the original LSH
paper~\cite{indyk98}, which is geared towards
\emph{distance} measures.

Locality-sensitive hashing schemes have been proposed for a
variety of similarity functions thus far, including Jaccard
similarity~\cite{broder98, li10}, Cosine
similarity~\cite{charikar02} and kernelized similarity functions
(representing e.g. a learned similarity metric)~\cite{jain08}.

\smallskip
\noindent \textbf{Candidate generation via LSH:}\\
One of the main reasons for the popularity of LSH is that it
can be used to construct an index that enables
efficient candidate generation for the similarity
search problem. Such LSH-based indices have been found to significantly
outperform more traditional indexing methods based on space
partitioning approaches, especially with increasing
dimensions~\cite{indyk98,datar04}. 
The general method works as 
follows~\cite{indyk98,datar04,broder97,ravichandran05,henzinger06}.
For
each object in the dataset, we
will form $l$ signatures, where each signature is
a concatentation of $k$ hashes. All pairs of objects that share
at least one of the $l$ signatures will be generated as a
candidate pair. Retrieving each pair of objects that share a
signature can be done efficiently using hashtables. 
For a given $k$ and similarity threshold $t$, the number of length-$k$
signatures required for an expected false negative rate $\epsilon$ 
can be shown to be $l =
\lceil \frac{\log \epsilon}{\log (1-t^k)} \rceil$~\cite{xiao11}. 

\smallskip
\noindent \textbf{Candidate verification and similarity
estimation:}\\
The similarity between the generated candidates can be computed
in one of two ways:(a) by exact calculation of the similarity
between each pair, or (b) using an estimate of the similarity, as
the fraction of hashes that the two objects agree upon. The pairs
of objects with estimated similarity greater than the threshold
are finally output. In terms of running time, approach (b) is
often faster, especially when the number of candidates is large
and/or exact similarity calculations are expensive, such as with
more complex similarity measures or with larger vector lengths. 
The main overhead with approach (b) is in hashing each point
sufficient number of times in the first place, but this cost is
amortized over many similarity computations (especially in the
case of all-pairs similarity search), and furthermore we
need the hashes for candidate generation in any case. However,
what is less clear is how good this simple estimation procedure
is in terms of accuracy, and whether it can be made any faster.
We will address these questions next.

\section{Classical similarity estimation for LSH}
\label{sec:mlest}

Similarity estimation for a candidate pair using LSH can 
be considered as a
statistical parameter inference problem. The parameter we wish
to infer is the similarity, and the data we observe is the
outcome of the comparison of each successive hash between the
candidate pair. The probability model relating the parameter to
the data is given by the main LSH equation,
Equation~\ref{eqn:lshmain}. There are two main schools of
statistical inference - classical (frequentist) and Bayesian. 

Under classical (frequentist)
statistical inference, 
the parameters of a probability model are treated as
fixed, and it is considered meaningless to make 
probabilistic statements about the parameters - hence the output
of classical inference is simply a point estimate, one for each
parameter. The best known example
of frequentist inference is maximum
likelihood estimation, where the value of the parameter that 
maximizes the
probability of the observed data is output as the point estimate. In the
case of similarity estimation via LSH, let us say we have
compared $n$ hashes and have observed $m$ agreements in hash
values. The maximum likelihood estimator for the similarity
$\hat{s}$ is:\footnote{Proofs are elementary and are omitted.}
\[
\hat{s} = \frac{m}{n}
\] 
While previous researchers have not explicitly labeled their
approaches as using the maximum likelihood estimators, they have
implicitly used the above estimator, tuning the number of hashes
$n$~\cite{ravichandran05,charikar02}. However, this approach has
some important drawbacks, which we turn to next.

\subsection{Difficulty of tuning the number of hashes}
\label{subsub:manyhashes}

While the above estimator is unbiased, the variance is 
$\frac{s * (1-s)}{n}$, meaning that the variance of the
estimator
depends on the similarity $s$ being estimated. 
This indicates that in
order to get the same level of accuracy for different
similarities, we will need to use \emph{different} number of
hashes.

\begin{wrapfigure}{r}{100pt}
\vspace{-0.25in}
\begin{center}
\hspace{-0.25in}
\includegraphics[width=110pt,height=110pt]{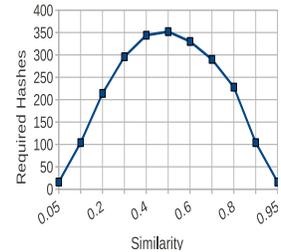}
\vspace{-0.3in}
\caption{Hashes vs. similarity}
\label{fig:hashes}
\end{center}
\vspace{-0.2in}
\end{wrapfigure}

We can be more precise and, for a given similarity, 
calculate exactly the the probability of a
smaller-than-$\delta$ error in $\hat{s}_{n}$, the similarity 
estimated using $n$ hashes. 
\begin{eqnarray*}
Pr[ |\hat{s}_{n} - s| < \delta] &=& Pr[ (s-\delta)*n \leq m \leq
(s+\delta)*n ] \\
&=& \sum_{m=(s-\delta)*n}^{(s+\delta)*n} {n \choose m} s^m
(1-s)^{n-m}
\end{eqnarray*}
Using the above expression, we can calculate the minimum number
of hashes needed to ensure that the similarity estimate is
sufficiently concentrated, i.e  within $\delta$ of the true value
with probability $1-\gamma$. A plot of the number of hashes required for
$\delta=\gamma=0.05$ for various similarity values is given in
Figure~\ref{fig:hashes}. As can be seen, there is a great
difference in the number of hashes required when the true
similarities are different; similarities closer to 0.5 require
far more hashes to estimate accurately than similarities close to
0 or 1. A similarity of 0.5 needs 350 hashes for
sufficient accuracy, but a similarity of 0.95 needs only 16
hashes!
Stricter accuracy
requirements lead to even greater differences in the required
number of hashes. 

Since we don't know the true similarity of each pair a priori, we cannot
choose the right number of hashes beforehand. 
If we err on the side of accuracy and choose a
large $n$, then performance suffers since we will be
comparing many more hashes than are necessary for some candidate
pairs. If, on the other hand, we err on the side of performance
and choose a smaller $n$, then accuracy suffers. With standard
similarity estimation, therefore, it is \emph{impossible} to tune the
number of hashes for the entire dataset so as to achieve both
optimal performance and accuracy.

\subsection{Ignores the potential for early pruning}
\label{subsub:noprune}

In the context of similarity search with a user-specified
threshold, the standard similarity estimation procedure also
misses opportunities for \emph{early candidate pruning}. The
intuition here is best illustrated using an example: Let us say the
similarity threshold is $0.8$ i.e. the user is only interested in
pairs with similarity greater than $0.8$. Let us say the
similarity estimation is going to use $n=1000$ hashes. But if we
are examining a candidate pair for which, out of the first $100$
hashes, only $10$ hashes matched, then intuitively it seems very
likely that this pair does not meet the threshold of $0.8$. In
general, it seems intuitively possible to be able to prune away
many false positive candidates by looking only at the first few
hashes, without needing to compare all the hashes. 
As we will see, most candidate generation algorithms produce
significant number of false positives, and the standard
similarity estimation procedure using LSH does not exploit the
potential for early pruning of candidate pairs.

\section{Candidate pruning and similarity estimation using
BayesLSH}

The key characteristic of Bayesian statistics is that it allows
one to make probabilistic statements
about any aspect of the world,
including things that would be considered ``fixed'' under
frequentist statistics and hence meaningless to make
probabilistic statements about. 
In particular, Bayesian statistics allows us to make
probabilistic statements about the parameters of probability models
- in other words, parameters are also treated as 
random variables. Bayesian inference generally consists of
starting with a \emph{prior} distribution over the parameters,
and then computing a \emph{posterior} distribution over the
parameters, conditional on the data that we have actually
observed, using Bayes' rule. A commonly cited drawback of
Bayesian inference is the need for the prior probability
distribution over the parameters, but a reasonable amount of data
generally ``swamps out'' the influence of the prior (see
Appendix). Furthermore,
a good prior can often lead to \emph{improved} estimates over
maximum likelihood estimation - this is a common strategy for
avoiding overfitting the data in machine learning and
statistics. The big advantage of Bayesian inference in the context of
similarity estimation is that instead of just outputing a point
estimate of the similarity, it gives us the complete posterior
distribution of the similarity. 
In the rest of this section, we will avoid discussing
specific choices for the prior distribution and similarity
measure in order to keep the
discussion general.

Fix attention on a particular pair $(x,y)$, and let us say that
$m$ out of the first $n$ hashes match for this pair. We will
denote this event as $M(m,n)$. The
conditional probability of the event $M(m,n)$ given the
similarity $S$ (here $S$ is a random variable), is given by the
binomial distribution with $n$ trials, where the success of each
trial is $S$ itself, from the Equation~\ref{eqn:lshmain}. 
Note that we have already observed
the event $M(m,n)$ happening i.e. $m$ and $n$ are not random
variables, they are the data.
\begin{eqnarray}
Pr[ M(m,n)  \, | \,  S ] &=& {n \choose m} S^{m} {(1-S)}^{n-m}
\label{eqn:likelihood}
\end{eqnarray}
What we are interested in knowing is the probability distribution
of the similarity $S$, \emph{given} that we already know that
$m$ out of $n$ hashes have matched. Using Bayes' rule, the
posterior distribution for $S$ can be written as
follows:~\footnote{In terms of notation, we will use lower-case
$p(.)$ for probability density functions of continuous
random-variables. $Pr[.]$ is used for probabilities of discrete
events or discrete random variables.}
\begin{eqnarray*}
p( S  \, | \,  M(m,n) ) &=& \frac{ p( M(m,n)  \, | \,  S ) p(S)}{ p(M(m,n)) } \\
		&=& \frac{ p( M(m,n)  \, | \,  S ) p(S)}{ \int_0^1 {p(M(m,n), s) ds}} \\
		&=& \frac{ p( M(m,n)  \, | \,  S ) p(S)}{ \int_0^1 {p(M(m,n)  \, | \,  s)
		p(s) ds} }
\label{eqn:posterior}
\end{eqnarray*}
By plugging in the expressions for $p(M(m,n)  \, | \,  S)$ from
Equation~\ref{eqn:likelihood} and a suitable prior distribution
$p(S)$, we can get, for every value of $n$ and $m$, the posterior
distribution of $S$ conditional on the event $M(m,n)$. 
We calculate the following quantities in terms of the posterior
distribution:
\begin{enumerate}
\item
If after comparing $n$ hashes, $m$ matches agree, what is the
probability that the similarity is greater than the threshold $t$?
\begin{eqnarray}
Pr[ S \geq t  \, | \,  M(m,n) ] &=& \int_t^1 p(s  \, | \,  M(m,n)) ds 
\label{eqn:pruneProb}
\end{eqnarray}
\item
If after comparing $n$ hashes, $m$ matches agree, what is the
maximum-a-posteriori estimate for the similarity i.e. the
similarity value with the highest posterior probability? This
will function as our estimate $\hat{S}$ 
\begin{equation}
\hat{S} = \arg\max_{s} p(s | M(m,n))
\label{eqn:map}
\end{equation}
\item
Assume after comparing $n$ hashes, $m$ matches agree, and we
have estimated the similarity to be $\hat{S}$ (e.g. as indicated above). 
What is the concentration probability of $\hat{S}$ i.e. 
probability that this estimate is within $\delta$ of the true
similarity?
\begin{small}
\begin{eqnarray}
\hspace{-0.3in}
Pr[ |S - \hat{S}| < \delta \, | \, M(m,n) ] &=& Pr[ \hat{S}-\delta <
S < \hat{S}+\delta  \, | \,  M(m,n) ] \\
&=& \int_{\hat{S}-\delta}^{\hat{S}+\delta} p(s  \, | \,  M(m,n)) ds 
\label{eqn:concentrateProb}
\end{eqnarray}
\end{small}
\end{enumerate}
Assuming we can perform the above three kinds of inference, we
design our algorithm, BayesLSH, so that it satisfies the
probabilistic guarantees outlined in Section~\ref{sec:intro}. The
algorithm is outlined in Algorithm~\ref{alg:bayeslsh}.
For each candidate pair
$(x,y)$ we incrementally compare their respective hashes (line
8, the parameter $k$ indicates the number of hashes we
will compare at a time), until either one of two events happens. 
The first possibility
is that the candidate pair
gets pruned away because the probability of it being a true
positive pair has become very small (lines 10, 11 and 12), where
we use Equation~\ref{eqn:pruneProb} to calculate this
probability. The alternative possibility is that the candidate
pair does not get pruned away, and we continue comparing hashes
until our similarity estimate (line 14) becomes sufficiently
concentrated that it passes our accuracy requirements (lines 15
and 16). Here we use Equation~\ref{eqn:concentrateProb} to
determine the probability that our estimate is sufficiently
accurate. Each such pair is added to the output set of candidate
pairs, along with our similarity estimate (lines 19 and 20). 

Our second algorithm, BayesLSH-Lite (see
Algorithm~\ref{alg:lite}) is a simpler version of
BayesLSH, which calculates similarities exactly. Since the similarity
calculations are exact, there is no need for parameters $\delta,
\gamma$; however, this comes at the cost of some intuitiveness,
as there is a new parameter $h$ specifying the
maximum number of hashes that will be examined for each pair of
objects. BayesLSH-Lite
can be faster than BayesLSH for those datasets where exact
similarity calculations are cheap, e.g. because the object
representations are simpler, such as binary, or if 
the average size of the objects is small. 

BayesLSH clearly overcomes the two drawbacks of standard
similarity estimation explained in
Sections~\ref{subsub:manyhashes} and \ref{subsub:noprune}. Any
candidate pairs that can be pruned away by examining only the
first few hashes will be pruned away by BayesLSH. 
As we will show later, this
method is very effective for pruning away the vast majority of
false positives. Secondly, the number of hashes for which each
candidate pair is compared is determined automatically by the
algorithm, depending on the user-specified accuracy requirements,
completely eliminating the need to manually set the number of
hashes. Thirdly, each point in the dataset is only hashed as many
times as is necessary. This will be particularly useful for
applications where hashing a point itself can be costly e.g. for
kernel LSH~\cite{jain08}. Also, outlying points which don't have any
points with whom their similarity exceeds the threshold need only
be hashed a few times before BayesLSH prunes all candidate pairs
involving such points away. 

\algsetup{indent=2em}
\begin{algorithm}
\caption{BayesLSH}
\begin{small}
\begin{algorithmic}[1]
\label{alg:bayeslsh}

\STATE \textbf{Input:} Set of candidate pairs $C$; Similarity
threshold $t$; recall parameter $\epsilon$; accuracy parameters $\delta, \gamma$
\STATE \textbf{Output:} Set $O$ of pairs $(x,y)$ along with similarity estimates
$\hat{S}_{x,y}$ 

\smallskip

\STATE $O \leftarrow \emptyset$

\smallskip

\FORALL{$(x,y) \in C$}
	\STATE	$n,m \leftarrow 0$ \COMMENT{Initialization}
	\STATE $isPruned \leftarrow$ False

	\smallskip
	
	\WHILE{True}
		\STATE $m = m + \sum_{i=n}^{n+k} I[h_{i}(x) == h_{i}(y)]$
		\COMMENT{Compare hashes $n$ to $n+k$}
		\STATE $n = n + k$

		\IF{ $Pr[ S \geq t \, | \, M(m,n) ] < \epsilon$ }
			\STATE $isPruned \leftarrow$ True
			\STATE \textbf{break}
			\COMMENT{Prune candidate pair}
		\ENDIF

		\STATE $\hat{S} \leftarrow \arg\max_{s} p( s | M(m,n) )$

		\IF{ $Pr[ |S - \hat{S}\, | \, M(m,n) < \delta] < \gamma$}
			\STATE \textbf{break}
			\COMMENT{Similarity estimate is sufficiently concentrated}
		\ENDIF

	\ENDWHILE

	\smallskip

	\IF{ $isPruned$ == False }
	\STATE $O \leftarrow O \cup \{((x,y), \hat{S})\}$
	\ENDIF
\ENDFOR

\RETURN $O$

\end{algorithmic}
\end{small}
\end{algorithm}

\algsetup{indent=2em}
\begin{algorithm}
\caption{BayesLSH-Lite}
\begin{small}
\begin{algorithmic}[1]
\label{alg:lite}

\STATE \textbf{Input:} Set of candidate pairs $C$; Similarity
threshold $t$; recall parameter $\epsilon$; Number of hashes to use $h$
\STATE \textbf{Output:} Set $O$ of pairs $(x,y)$ along with exact
similarities $\hat{S}_{x,y}$ 

\smallskip

\STATE $O \leftarrow \emptyset$

\smallskip

\FORALL{$(x,y) \in C$}
	\STATE	$n,m \leftarrow 0$ \COMMENT{Initialization}
	\STATE $isPruned \leftarrow$ False

	\smallskip
	
	\WHILE{$n < h$}
		\STATE $m = m + \sum_{i=n}^{n+k} I[h_{i}(x) == h_{i}(y)]$
		\COMMENT{Compare hashes $n$ to $n+k$}
		\STATE $n = n + k$

		\IF{ $Pr[ S \geq t \, | \, M(m,n) ] < \epsilon$ }
			\STATE $isPruned \leftarrow$ True
			\STATE \textbf{break}
			\COMMENT{Prune candidate pair}
		\ENDIF
	\ENDWHILE

	\smallskip

	\IF{ $isPruned$ == False }
		\STATE $s_{x,y} = similarity(x,y)$
		\COMMENT{Exact similarity}
		\IF{ $s_{x,y} > t$ }
			\STATE $O \leftarrow O \cup \{((x,y), s_{x,y})\}$
		\ENDIF
	\ENDIF
\ENDFOR

\RETURN $O$

\end{algorithmic}
\end{small}
\end{algorithm}

In order to obtain a concrete instantiation of BayesLSH, we will
need to specify three aspects: (i) the LSH family of hash functions,
(ii) the choice of prior and (iii) how to tractably perform
inference. Next, we will look at specific instantiations of BayesLSH for
different similarity measures.

\subsection{BayesLSH for Jaccard similarity}

We will first discuss how BayesLSH can be used for approximate
similarity search for Jaccard similarity.

\smallskip
\noindent \textbf{LSH family:}
The LSH family for Jaccard similarity is the family of minwise
independent permutations~\cite{broder98, bohman00} on the universe from which our
collection of sets is drawn. Each hash function
returns the minimum element of the input set when the elements of the
set are permuted as specified by the hash function (which itself
is chosen at random from the family of minwise independent
permutations). The output of this family of hash functions,
therefore, is an integer representing the minimum element of the
permuted set. 

\smallskip
\noindent \textbf{Choice of prior:}
It is common practice in Bayesian inference to choose priors from
a family of distributions that is \emph{conjugate} to the 
likelihood distribution, so that the inference is tractable and
also that the posterior belongs to the same distribution family
as the prior (indeed, that is the definition of a conjugate
prior). 
The likelihood in this case is given by a binomial
distribution, as indicated in Equation~\ref{eqn:likelihood}. The
conjugate for the binomial is the \emph{Beta} distribution, which
has two parameters $\alpha > 0, \beta
> 0$ and is defined on the domain $(0,1)$. The
pdf for $Beta(\alpha, \beta)$ is defined as follows.
\[
p( s ) = \frac{s^{\alpha-1}*(1-s)^{\beta-1}}{B(\alpha, \beta)}
\]
Here $B(\alpha, \beta)$ is the beta function, and it can also be
thought of as a normalization constant to ensure the entire
distribution integrates to $1$.

Even assuming we want to model the prior using a Beta
distribution, how do we choose the parameters $\alpha, \beta$? A
simple choice is to set $\alpha=1, \beta=1$, which results in a
uniform distribution on $(0,1)$. However, we can actually
learn $\alpha, \beta$ so as to best fit a random sample of
similarities from candidate pairs output by the candidate
generation algorithm.
Let us assume we have $r$ samples chosen
uniformly at random from the total population of candidate pairs generated
by the particular candidate generation algorithm being used, and
their similarities are $s_1, s_2, \ldots, s_r$. Then we can
estimate $\alpha, \beta$ so as to best model the distribution of
similarities among candidate pairs. For Beta distribution, a
simple and effective method of learning the parameters is via 
method-of-moments estimation. In this method, we calculate the
sample moments (sample mean and sample variance), assume that
they are the true moments of the distribution and solve for the
parameter values that will result in the obtained moments. In our
case, we have the following estimates for $\alpha, \beta$:
\[
\hat{\alpha} = \bar{s} \left( \frac{\bar{s}(1-\bar{s})}{\bar{s}_{v}} -
1 \right) \; ; \; \hat{\beta} = (1-\bar{s}) \left( \frac{\bar{s}(1-\bar{s})}{\bar{s}_{v}} -
1 \right)
\]
where $\bar{s}$ and ${\bar{s}_v}$ are the sample mean and
variance, given as follows:
\[
\bar{s} = \frac{ \sum_{i=1}^{r}{s_i}}{r} \; ; \; \bar{s}_{v} = \frac{ \sum_{i=1}^{r}{{(s_i-\bar{s})}^2}}{r}
\]

Assuming a prior $Beta(\alpha, \beta)$ distribution on the
similarity, and we observe the event $M(m,n)$ i.e. $m$ out of the
first $n$ hashes match, then the posterior distribution of the
similarity looks as follows:
\begin{eqnarray*}
p(s | M(m,n)) &=& \frac{{n \choose m} s^{m} {(1-s)}^{n-m} s^{\alpha-1}
{(1-s)}^{\beta-1}}
{\int_0^1{{n \choose m} s^{m} {(1-s)}^{n-m} s^{\alpha-1}
{(1-s)}^{\beta-1}}} \\
&=& \frac{s^{m+\alpha-1} {(1-s)}^{n-m+\beta-1}}{B(m+\alpha,
n-m+\beta)}
\end{eqnarray*}

Hence, the posterior distribution of the similarity also follows
a Beta distribution with parameters $m+\alpha$ and $n-m+\beta$.

\smallskip
\noindent \textbf{Inference:} We next show concrete
ways to perform inference, i.e. computing
Equations~\ref{eqn:pruneProb}, \ref{eqn:map} and \ref{eqn:concentrateProb}.

The probability that similarity is greater than the threshold
after observing that $m$ out of the first $n$ hashes match is:
\begin{eqnarray*}
Pr [ S \geq t | M(m,n) ] &=& \int_t^1 p( s | M(m,n) ) \\
&=& 1 - I_t(m+\alpha, n-m+\beta)
\end{eqnarray*}
Above $I_t(.,.)$ refers to the \emph{regularized incomplete beta}
function, which gives the cdf for the beta distribution. This function is available in standard scientific 
computing libraries, where it is typically approximated 
using continued fractions~\cite{didonato92}.

Our similarity estimate, after observing $m$ matches in $n$
hashes, will be the mode of the posterior distribution $p(s |
M(m,n))$. The mode of $Beta(\alpha, \beta)$ is given by
$\frac{\alpha-1}{\alpha+\beta-2}$. Therefore, our similarity
estimate after observing that $m$ out of the first $n$ hashes
agree is $\hat{S} =
\frac{m+\alpha-1}{n+\alpha+\beta-1}$.

The concentration probability of the similarity estimate
$\hat{S}$ can be derived as follows (the expression for $\hat{S}$
indicated above can be substituted in the below equations):
\begin{small}
\begin{eqnarray*}
Pr [ |\hat{S}-S| < \delta | M(m,n) ] &=&
\int_{\hat{S}-\delta}^{\hat{S}+\delta} p( s | M(m,n) ) ds \\
&=&
I_{\hat{S}+\delta}(m+\alpha, n-m+\beta) \\
&-& I_{\hat{S}-\delta}(m+\alpha, n-m+\beta) 
\end{eqnarray*}
\end{small}


Thus by substituting the above computations in the corresponding
places in Algorithm~\ref{alg:bayeslsh}, we obtain a version of
BayesLSH specifically adapted to Jaccard similarity.

\subsection{BayesLSH for Cosine similarity}

We will next discuss instantiating BayesLSH for Cosine
similarity.

\smallskip
\noindent \textbf{LSH family:}
For Cosine similarity, each hash function $h_{i}$
is associated with a random vector $r_{i}$, each of whose
components is a sample from the standard gaussian ($\mu=0,
\sigma=1$). For a vector $x$, $h_i(x) = 1$ if $dot(r_i,x) \geq 0$
and $h_i(x)=0$ otherwise~\cite{charikar02}. Note that each hash
function outputs a bit, and hence these hashes can be stored
with less space.

However, there is one challenge here that needs to be overcome
that was absent for BayesLSH with Jaccard similarity: 
this LSH family is for a slightly
different similarity measure than cosine - it is instead for $1 -
\frac{\theta(x,y)}{\pi}$, where $\theta(x,y) =
\arccos{(\frac{dot(x,y)}{||x||.||y||})}$. For notational ease, we will refer to this similarity function as
$r(x,y)$ i.e. $r(x,y) = 1 - \frac{\theta(x,y)}{\pi}$. Explicitly,
\[
Pr[ h_i(x) == h_i(y) ] = r(x,y)
\]
\[
Pr[ M(m,n) | r ] = {n \choose m} r^m (1-r)^{n-m}
\]
Since the similarity function we are interested in is $cos(x,y)$
and not $r(x,y)$ - in particular, we wish for probabilistic
guarantees on the quality of the output in terms of $cos(x,y)$
and not $r(x,y)$ - we will need to somehow express the posterior
probability in terms of $s = cos(x,y)$.
One can choose to re-express the likelihood in terms of $s =
cos(x,y)$ instead of in terms of $r$ but this introduces $cos()$
terms into the likelihood, and makes it very hard to find a
suitable prior that keeps the inference tractable. Instead we
compute the posterior distribution of $r$  which we transform
appropriately into a posterior distribution of $s$. 

\smallskip
\noindent \textbf{Choice of prior:}
We will need to choose a prior distribution for $r$.
Previously, we used a Beta prior for Jaccard BayesLSH;
unfortunately $r$ has range $[0.5,1]$, while the standard Beta
distribution has support on the domain $(0,1)$. We can still map
the standard Beta distribution onto the domain $(0.5,1)$, but
this distribution will no longer be conjugate to the binomial
likelihood.\footnote{The pdf of a Beta distribution supported
only on $(0.5,1)$ with parameters $\alpha, \beta$ is $p(x)
\propto {(x-0.5)}^{\alpha-1} {(1-x)}^{\beta-1}$. With a binomial
likelihood, the posterior pdf takes the form $p( x | M(m,n) ) \propto
x^m {(x-0.5)}^{\alpha-1} {(1-x)}^{n-m+\beta-1}$. Unfortunately there
is no simple and fast way to integrate this pdf.}
Our solution is to use a simple uniform distribution
on $[0.5,1]$ as the prior for $t$.Even when the true
similarity distribution is very far from being uniform 
(as is the
case in real datasets, including the ones used in our
experiments), this
prior still works well because the posterior is strongly
influenced by the actual outcomes observed (see Appendix).

The prior pdf therefore is:
\[
p(r) = \frac{1}{1-0.5} = 2
\]

The posterior pdf, after observing that $m$ out of the first $n$
hashes agree, is:
\begin{small}
\begin{eqnarray*}
p( r | M(m,n) ) &=& \frac{ 2 {n \choose m} r^m (1-r)^{n-m}}
		{\int_{0.5}^1 2 {n \choose m} r^m (1-r)^{n-m} dr} \\
 &=& \frac{ r^m (1-r)^{n-m}}
		{\int_{0.5}^1 r^m (1-r)^{n-m} dr} \\
&=& \frac{ r^m (1-r)^{n-m}}{ B_{1}(m+1, n-m+1) - B_{0.5}(m+1,
n-m+1)}
\end{eqnarray*}
\end{small}
Here $B_{x}(a,b)$ is the incomplete Beta function, defined as
$B_{x}(a,b) = \int_0^x {y^{a-1}{(1-y)}^{b-1}dy}$. 

\smallskip
\noindent \textbf{Inference}:
In order to calculate Equations~\ref{eqn:pruneProb},
\ref{eqn:concentrateProb} and \ref{eqn:map}, we will first need a
way to convert from $r$ to $s$ and vice-versa. Let $r2c: [0.5,1]
\rightarrow [0,1]$ be the 1-to-1 function that maps from $r(x,y)$
to $cos(x,y)$; $r2c()$ is given by $r2c(r) = \cos(\pi*(1-r))$.
Similarly, let $c2r$ be the 1-to-1 function that does the same
map in reverse; $c2r()$ is given by $c2r(c) =
1-\frac{\arccos(c)}{\pi}$.

Let $R$ be the random variable such that $R=c2r(S)$ and let $t_r
= c2r(t)$. After observing that the $m$ out
of the first $n$ hashes agree, the probability that cosine similarity is greater 
than the threshold $t$ is:
\begin{small}
\begin{eqnarray*}
\hspace{-0.2in}
 Pr[ S \geq t | M(m,n) ]
&=& Pr [ c2r(S) \geq c2r(t) | M(m,n) ] \\
&=& Pr [ R \geq t_r | M(m,n) ] \\
&=& \int_{t_r}^1 p( r | M(m,n) ) dr \\
&=& \frac{\int_{t_r}^1 r^m (1-r)^{n-m} dr}
	{ B_{1}(m+1, n-m+1) - B_{0.5}(m+1,n-m+1)}\\
&=& \frac{B_{1}(m+1, n-m+1) -B_{t_r}(m+1,n-m+1)}{ B_{1}(m+1, n-m+1) - B_{0.5}(m+1,n-m+1)}\\
\end{eqnarray*}
\end{small}
The first step in the above derivation follows because $c2r()$ is a 
1-to-1 mapping. Thus, we have a concrete expression for
calculating Eqn~\ref{eqn:pruneProb}.

Next, we need an expression for the similarity estimate
$\hat{S}$, given that $m$ out of $n$ hashes have matched so far.
Let $\hat{R} = \arg\max_{r} p(r | M(m,n))$. We can obtain a
closed form expression for $\hat{R}$ by solving for
$\frac{\partial p(r | M(m,n) }{\partial r} = 0$; when we do this,
we get $r =
\frac{m}{n}$. Hence, $\hat{R} = \frac{m}{n}$. Now $\hat{S} =
r2c(\hat{R})$, therefore $\hat{S} = r2c(\frac{m}{n})$. This is
our expression for calculating Eqn~\ref{eqn:map}.

Next, let us consider the concentration probability of $\hat{S}$.
\begin{small}
\begin{eqnarray*}
&& Pr [ |\hat{S}-S| < \delta | M(m,n) ] \\
&=& Pr [ \hat{S}-\delta < S < \hat{S}+\delta | M(m,n) ] \\
&=& Pr [ c2r(\hat{S}-\delta) < c2r(S) < c2r(\hat{S}+\delta) |
M(m,n) ] \\
&=& Pr [ c2r(\hat{S}-\delta) < R < c2r(\hat{S}+\delta) |
M(m,n) ] \\
&=& \frac{\int_{c2r(\hat{S}-\delta)}^{c2r(\hat{S}+\delta)} r^m (1-r)^{n-m} dr}
	{ B_{1}(m+1, n-m+1) - B_{0.5}(m+1,n-m+1)} \\
&=& \frac{B_{c2r(\hat{S}+\delta)}(m+1,n-m+1) -
B_{c2r(\hat{S}-\delta)}(m+1,n-m+1)}{B_{1}(m+1, n-m+1) -
B_{0.5}(m+1,n-m+1)} 
\end{eqnarray*}
\end{small}

Thus, we have concrete expressions for
Equations~\ref{eqn:pruneProb}, \ref{eqn:map} and
\ref{eqn:concentrateProb}, giving us an instantiation of
BayesLSH adapted to Cosine similarity.

\subsection{Optimizations}
The basic BayesLSH can be optimized without affecting the
correctness of the algorithm in a few ways. The main idea behind
the optimizations here is to minimize the number of times
inference has to be performed, in particular the
Equations~\ref{eqn:pruneProb} and \ref{eqn:concentrateProb}.

\smallskip
\noindent \textbf{Pre-computation of minimum
matches:} 
We pre-compute the minimum number of matches a candidate pair needs
to have in order for $Pr [ S \geq t | M(m,n) ] > \epsilon$ to be
true, thus completely eliminating the need for any online
inference in line 10 of Algorithm~\ref{alg:bayeslsh}. 
For every value of $n$ that we will consider (upto some
maximum), we pre-compute the function $minMatches(n)$ defined as
follows:
\[
minMatches(n) = \arg\min_{m} Pr[ S \geq t | M(m,n) ] \geq
\epsilon
\]
This can be done via
binary search, since $Pr [ S \geq t | M(m,n) ]$ increases
monotonically with $m$ for a fixed $n$. Now, for each candidate
pair, we simply check if the actual number of matches for that pair
at every $n$ is at least $minMatches(n)$.
Note that we will not encounter every possible value of $n$ upto
the maximum - instead, since we compare $k$ hashes at a
time, we need to compute $minMatches()$ only once for all multiples
of $k$ upto the maximum.

\smallskip
\noindent \textbf{Cache results of inference:} 
We maintain a cache indexed by $(m,n)$ that indicates whether or
not the similarity estimate that is obtained after $m$ hashes out
of $n$ agree is sufficiently concentrated or not
(Equation~\ref{eqn:concentrateProb}). Note that for
each possible $n$, we only need to cache the results for $m \geq
minMatches(n)$, since lower values of $m$ are guaranteed to
result in pruning. Thus, in the vast majority of cases, we can
simply fetch the result of the inference from the cache instead
of having to perform it afresh. 

\smallskip
\noindent \textbf{Cheaper storage of hash functions:} 
For cosine similarity, storing the random gaussian vectors
corresponding to each hash function can
take up a fair amount of space. 
To reduce this storage requirement, we 
developed a scheme for storing each float using only 2 bytes, by
exploiting the fact that random gaussian samples from the standard 0-mean,
1-standard deviation gaussian lie well within a small interval
around 0. Let us assume that all of our samples will lie within
the interval (-8,8) (it is astronomically unlikely that a
sample from the standard gaussian lies outside this interval).
For any float $x \in (-8,8)$, it can be represented as a 2-byte
integer $x' = \lfloor (x+8) *\frac{2^{16}}{16} \rfloor$. 
The maximum error of
this scheme is $0.0001$ for any real number in $(-8,8)$.

\section{Experiments}
We experimentally evaluated the performance of BayesLSH and
BayesLSH-Lite on 6 real datasets with widely varying characteristics (see
Table~\ref{tab:datasets}). 
\begin{itemize}
\item
\vspace{-0.06in}
{\bf RCV1} is a text corpus of Reuters articles and is a popular
bechmarking corpus for text-categorization
research~\cite{lewis04}. We use the standard pre-processed version
of the dataset with word stemming and tf-idf weighting. 
\item
\vspace{-0.06in}
{\bf Wiki} datasets. We pre-processed the article dump of the
English Wikipedia\footnote{http://download.wikimedia.org} - Sep 2010 version - to produce both a text
corpus of Wiki articles as well as the directed graph of
hyperlinks between Wiki articles. Our pre-processing includes the
removal of stop-words, removal of insignificant articles, and
tf-idf weighting (for both the the text and the graph). Words occuring 
at least 20 times in the entire corpus are used as features, resulting in a
dimensionality of 344,352. The {\bf
WikiWords100K} dataset consists of text vectors with at least 500
non-zero features, of which there are 100,528. The {\bf
WikiWords500K} dataset consists of vectors with at least 200
non-zero features, of which there are 494,244. The {\bf
WikiLinks} dataset consists of the entire article-article graph
among \textasciitilde 1.8M articles, with Tf-Idf weighting. 
\item
\vspace{-0.06in}
{\bf Orkut} consists of a subset of the (undirected) friendship
network among nearly 3M Orkut users, made available
by~\cite{mislove07}. Each user is represented as a weighted
vector of their friends, with Tf-Idf weighting.
\item
\vspace{-0.06in}
{\bf Twitter} consists of the directed graph of
follower/followeee relationships among the subset of Twitter
users with at least 1,000 followers, first collected by Kwak et.
al.~\cite{kwak10}. Each user is represented as
a weighted vector of the users they follow, with Tf-Idf
weighting.
\end{itemize}
\vspace{-0.06in}

We note that all our datasets represent realistic applications
for all pairs similarity search. Similarity search on text
corpuses can be useful for clustering, semi-supervised learning,
near-duplicate detection etc., while similarity search on the graph
datasets can be useful for link prediction, friendship
recommendation and clustering.  Also, in our experiments
we primarily focus on similarity search for general real-valued
vectors using Cosine similarity, as opposed to similarity search
for binary vectors (i.e. sets). Our reasons are as follows:

\vspace{0.05in}
\noindent{\bf 1.}
Representations of objects as general real-valued vectors are
generally more powerful and lead to better similarity
assessments, Tf-Idf style representations being the 
classic
example here (see~\cite{satuluri11b} for 
another example from
graph mining). 

\vspace{0.05in}
\noindent{\bf 2.}
Similarity search is generally harder on real-valued vectors.
With binary vectors (sets), most similarity measures are
directly proportional to the
overlap between the two sets, and it is easier to obtain bounds
on the
overlap between two sets by inspecting only a few elements of
each set, since each element in the set can only 
contribute the same, fixed number (1) to the overlap. On the
other hand, with general real-valued vectors, different
elements/features have different weights (also, the same feature
may have different weights across different vectors), meaning that it
is harder to bound the similarity by inspecting only a few elements
of the vector.

\begin{table}
\begin{small}
\begin{tabular}{|c|c|c|c|c|}
\hline
{\bf
Dataset} & {\bf Vectors} & {\bf Dimensions} & {\bf Avg. len} &
{\bf Nnz}\\
\hline
RCV1 & 804,414 & 47,236 & 76 & 61e6\\
\hline
WikiWords100K & 100,528 & 344,352 & 786 & 79e6\\
\hline
WikiWords500K & 494,244 & 344,352 & 398 & 196e6\\
\hline
WikiLinks & 1,815,914 & 1,815,914 & 24 & 44e6\\
\hline
Orkut & 3,072,626 & 3,072,626 & 76 & 233e6\\
\hline
Twitter & 146,170 & 146,170 & 1369 & 200e6\\
\hline
\end{tabular}
\end{small}
\caption{Dataset details. Nnz stands for number of non-zeros.}
\label{tab:datasets}
\end{table}

\subsection{Experimental setup}
We compare the following methods for all-pairs similarity
search.

\noindent {\bf 1.} \textbf{AllPairs}~\cite{bayardo07} (AP) is one of the state-of-the-art
approaches for all-pairs similarity search, especially for cosine
similarity on real-valued vectors. AllPairs is an exact
algorithm. 


\noindent {\bf 2,3.} \textbf{AP+BayesLSH, AP+BayesLSH-Lite}: These are variants of
BayesLSH and BayesLSH-Lite where the input is the candidate set
generated by AllPairs.

\noindent {\bf 4,5.} \textbf{LSH, LSH Approx}: These are two variants of the standard
LSH approach for all pairs similarity search.
For both LSH and LSH Approx, candidate pairs are generated as
described in Section~\ref{sec:bg}
; for LSH, similarities are calculated
exactly, whereas for LSH Approx, similarities are instead
\emph{estimated} using the standard maximum
likelihood estimator, as described in Section~\ref{sec:mlest}. 
For LSH Approx, we tuned the number of hashes and set it 
to 2048 for cosine
similarity and 360 for Jaccard similarity. Note that the
hashes for Cosine similarity are only bits, while the hashes for
Jaccard are integers. 


\noindent {\bf 6,7.} \textbf{LSH+BayesLSH, LSH+BayesLSH-Lite}: 
These are variants of BayesLSH that take as input the candidate set 
generated by LSH as described in Section~\ref{sec:bg}.


\noindent {\bf 8.} \textbf{PPJoin+}~\cite{xiao08} is a state-of-the-art 
exact algorithm for
all-pairs similarity search, however it only works for binary
vectors and we only include it in the experiments with Jaccard
and binary cosine similarity.

For all BayesLSH variants, we report the full execution time i.e.
including the time for candidate generation. For BayesLSH variants,
$\epsilon=\gamma=0.03$ and $\delta=0.05$ ($\gamma, \delta$ don't
apply to BayesLSH-Lite). For the number of hashes to be compared
at a time, $k$, it makes sense to set this to be a multiple of
the word size, since for cosine similarity, each hash is simply a
bit. We set $k=32$, although higher multiples of the word size
work well too. 
In the case of BayesLSH-Lite, the number
of hashes to be used for pruning was set to $h=128$ for Cosine
and $h=64$ for Jaccard. For LSH and LSH
Approx, the expected false negative rate is set to $0.03$
. The randomized algorithms (LSH variants,
BayesLSH variants) were each run 3 times and the average results
are reported.

All of the methods work for both Cosine and Jaccard similarities,
for both real-valued as well as binary vectors, except for
PPJoin+, which only works for binary vectors. The code for
PPJoin+ was downloaded from the authors' website, all the
other methods were implemented by us.\footnote{Our AllPairs implementation
is slightly faster than the original implementation of the
authors due to a simple implementational fix. This has since been
incorporated into the authors' implementation.} All
algorithms are single-threaded and are implemented in C/C++. The
experiments were run by submitting jobs to a cluster, where each
node on the cluster runs on a dual-socket, dual-core 2.3 GHz Opteron
with 8GB RAM. Each algorithm was allowed 50 hrs (180K
secs) before it was declared timed out and killed.

We executed the different algorithms on both the weighted and
binary versions of the datasets, using Cosine similarity for the weighted
case and both Jaccard and Cosine for the binary case. For Cosine
similarity, we varied the similarity threshold from 0.5 to 0.9,
but for Jaccard we found that very few pairs satisfied higher
similarity thresholds (e.g. for Orkut, a 3M record dataset, only 1648 pairs were
returned at threshold 0.9), and hence varied the threshold from
0.3 to 0.7. For Jaccard and Binary Cosine, we only report results
on WikiWords500K, Orkut and Twitter, which are our three largest
datasets in terms of total number of non-zeros.

\subsection{Results comparing BayesLSH variants with baselines}
\begin{figure*}
\begin{center}
\includegraphics[width=320pt, height=40pt]{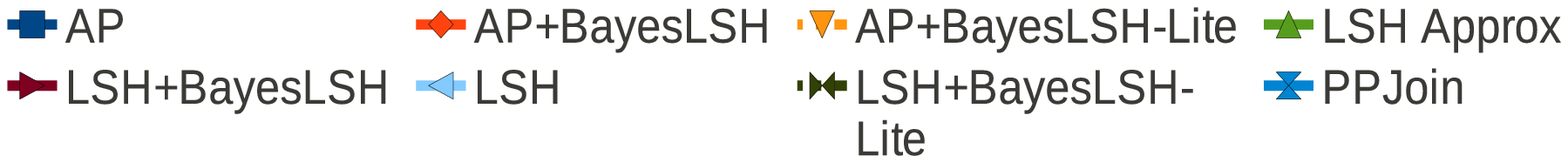}
\end{center}
\vspace{-0.33in}
\begin{center}
\subfigure[RCV1]
{
\includegraphics[width=160pt,height=135pt]{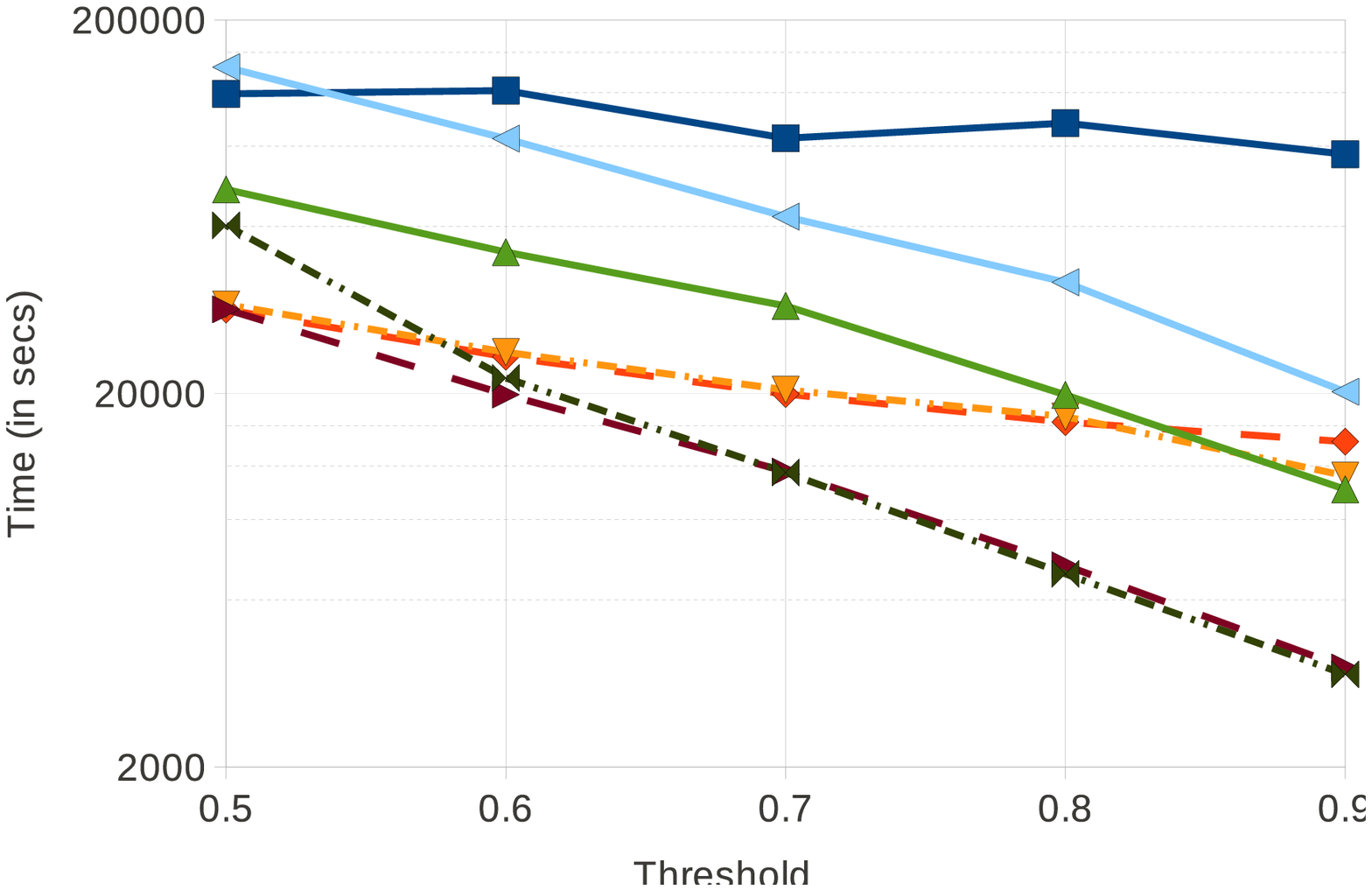}
\label{fig:rcv1}
}
\subfigure[WikiWords100K]
{
\includegraphics[width=160pt,height=135pt]{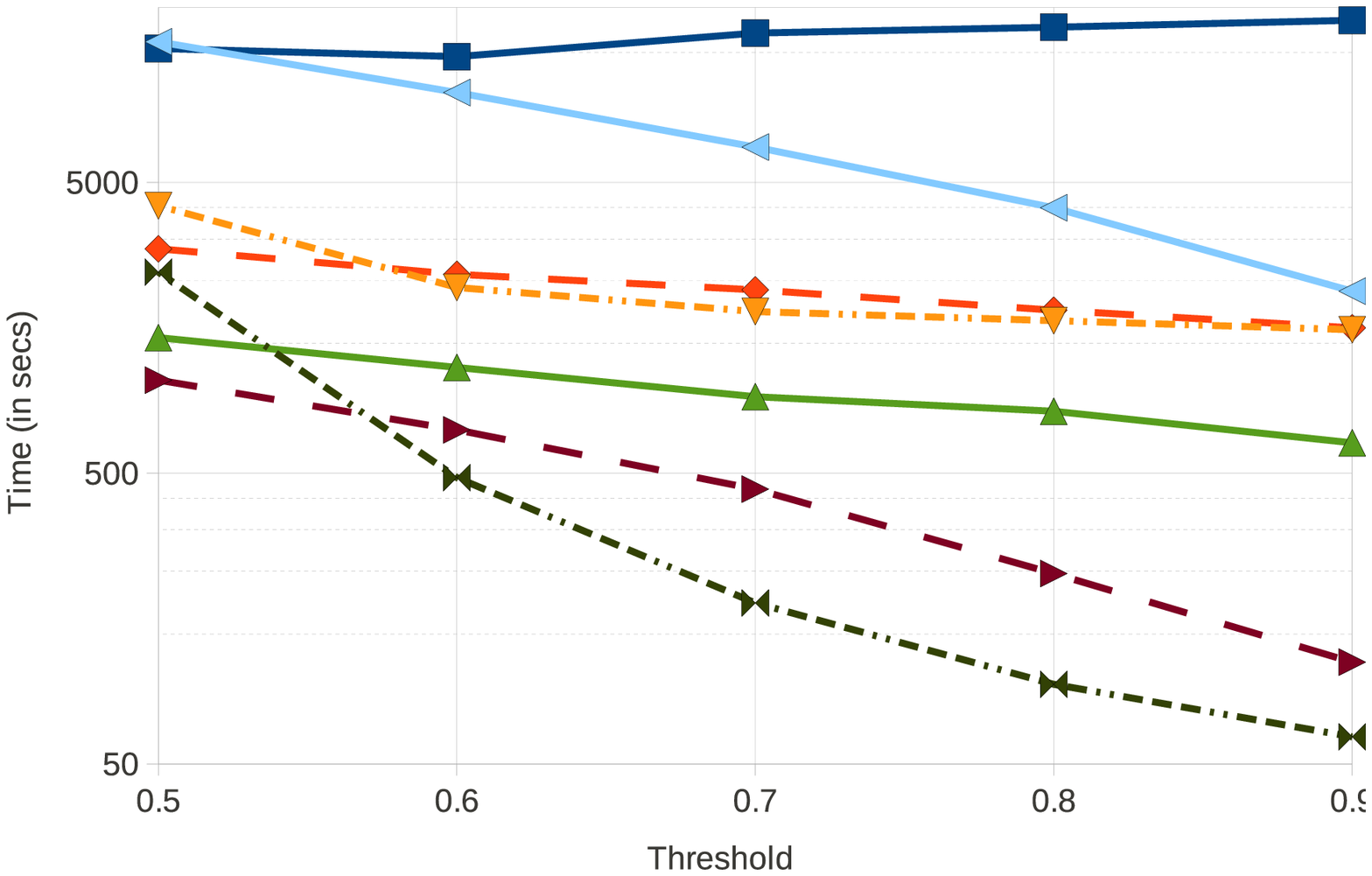}
\label{fig:wiki100k}
}
\subfigure[WikiWords500K]
{
\includegraphics[width=160pt,height=135pt]{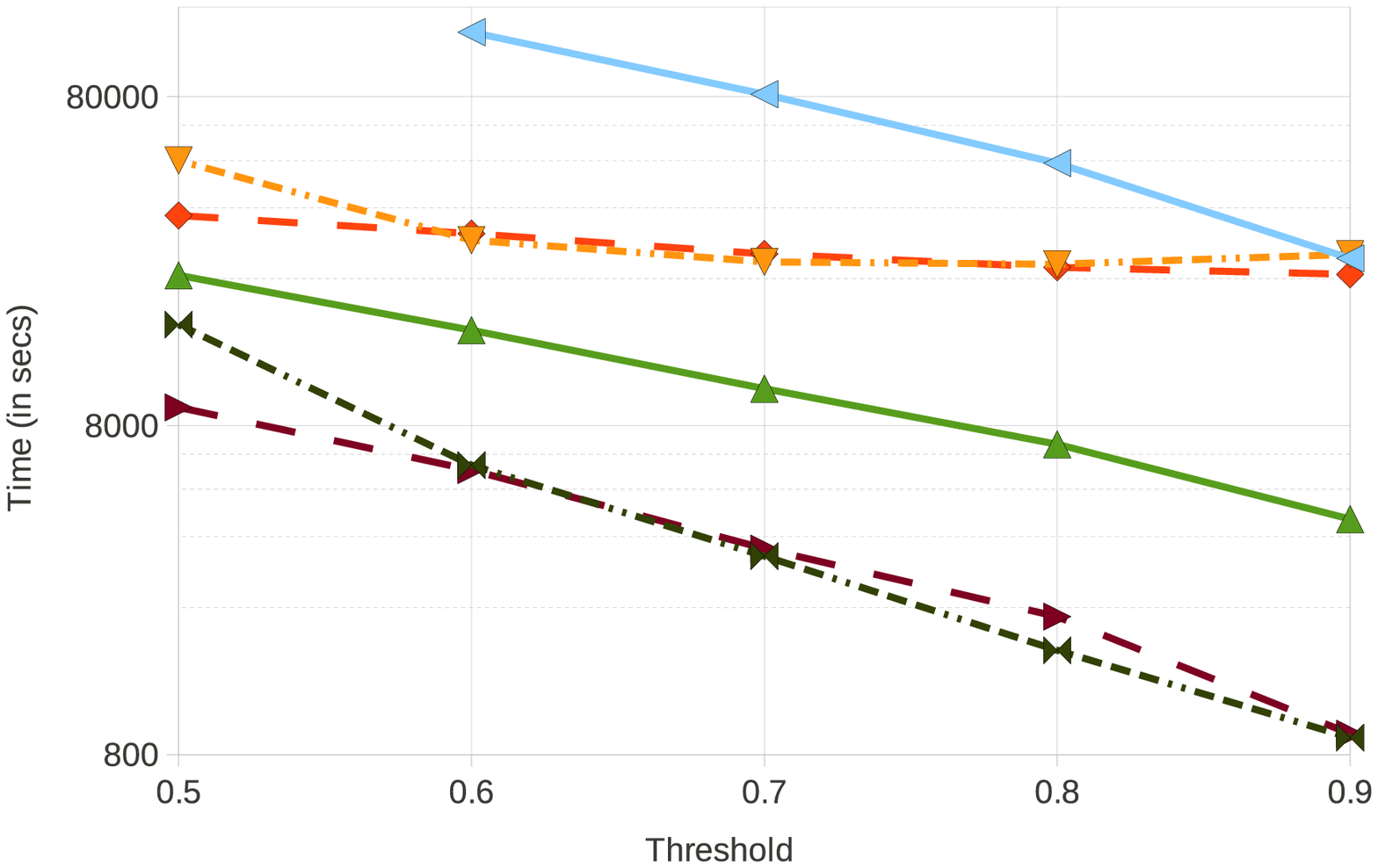}
\label{fig:wiki500k}
}
\end{center}
\vspace{-0.28in}
\begin{center}
\subfigure[WikiLinks]
{
\includegraphics[width=160pt,height=135pt]{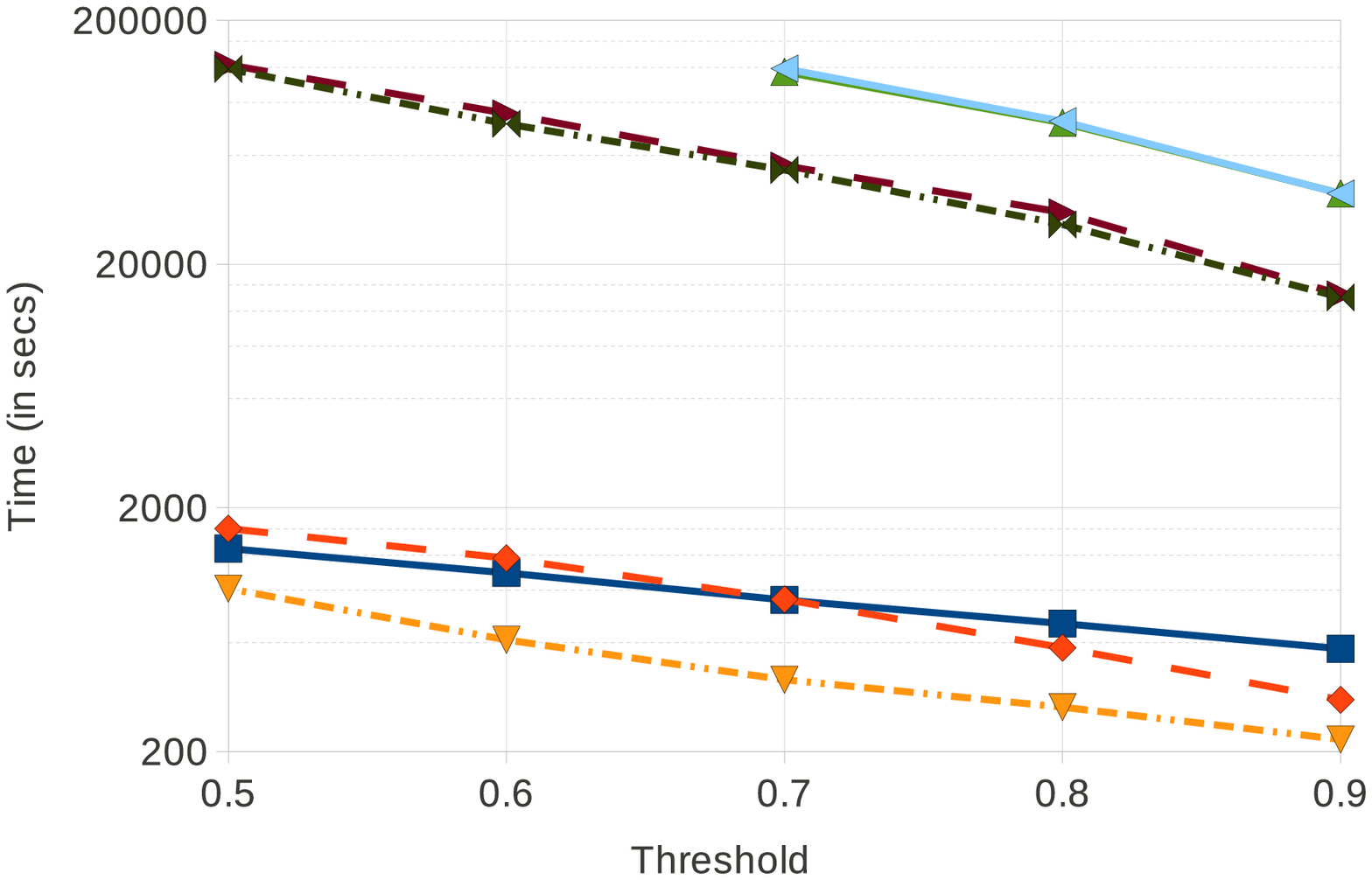}
\label{fig:wikiLinks}
}
\subfigure[Orkut]
{
\includegraphics[width=160pt,height=135pt]{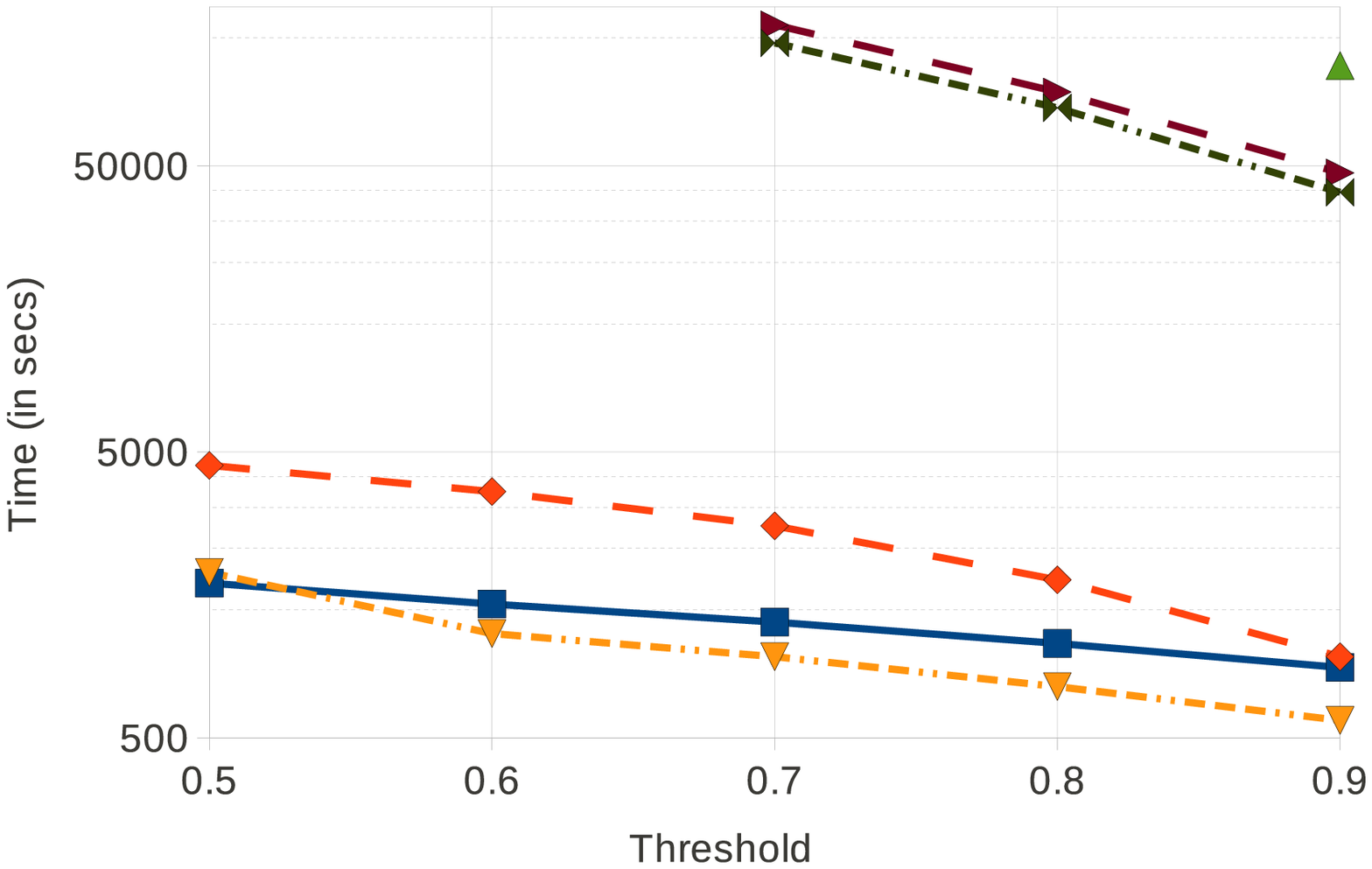}
\label{fig:orkut}
}
\subfigure[Twitter]
{
\includegraphics[width=160pt,height=135pt]{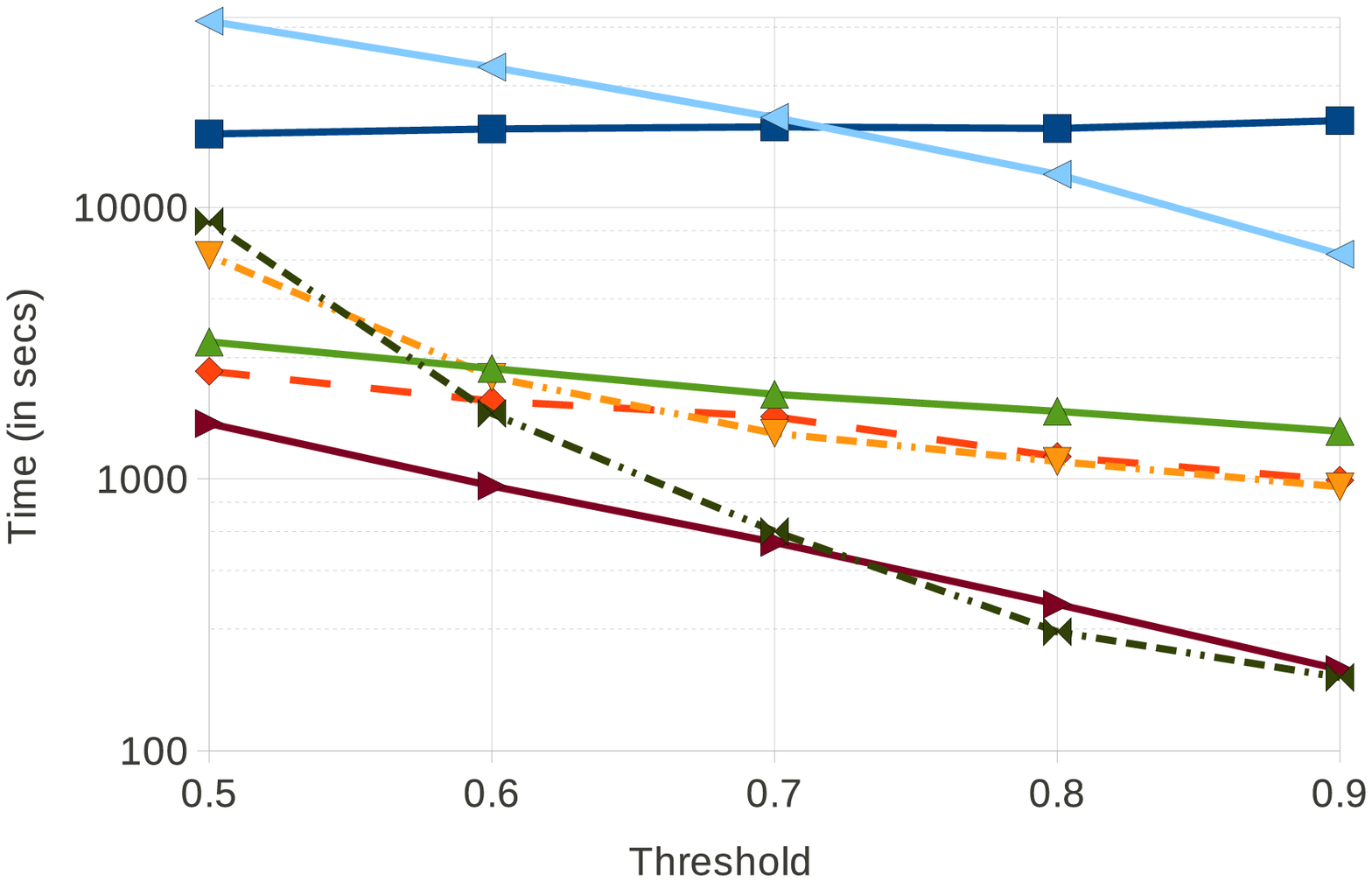}
\label{fig:twitter}
}
\end{center}
\vspace{-0.28in}
\begin{center}
\subfigure[WikiWords500K (Binary, Jaccard)]
{
\includegraphics[width=160pt,height=135pt]{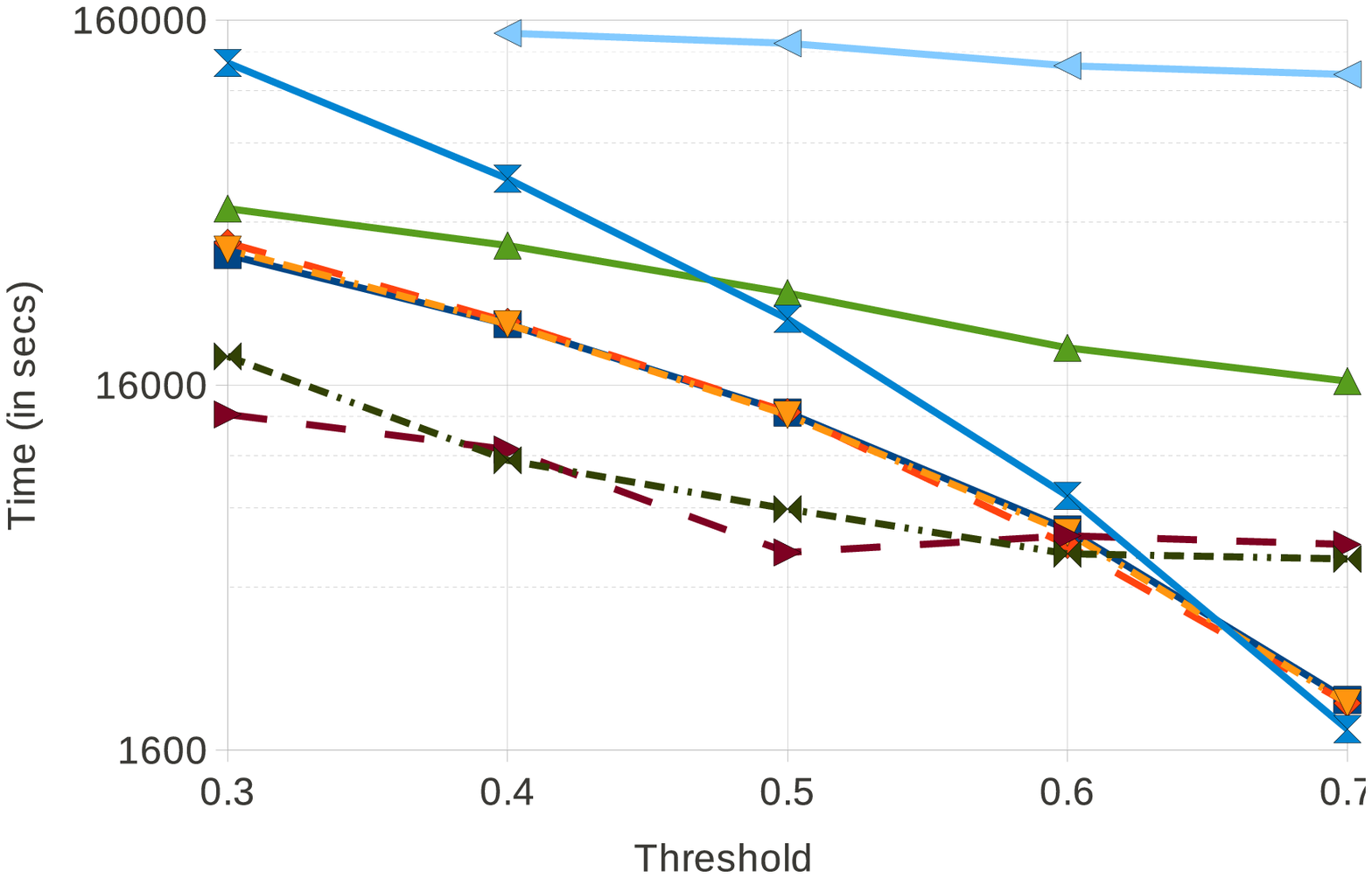}
\label{fig:wiki500K_jac}
}
\subfigure[Orkut (Binary, Jaccard)]
{
\includegraphics[width=160pt,height=135pt]{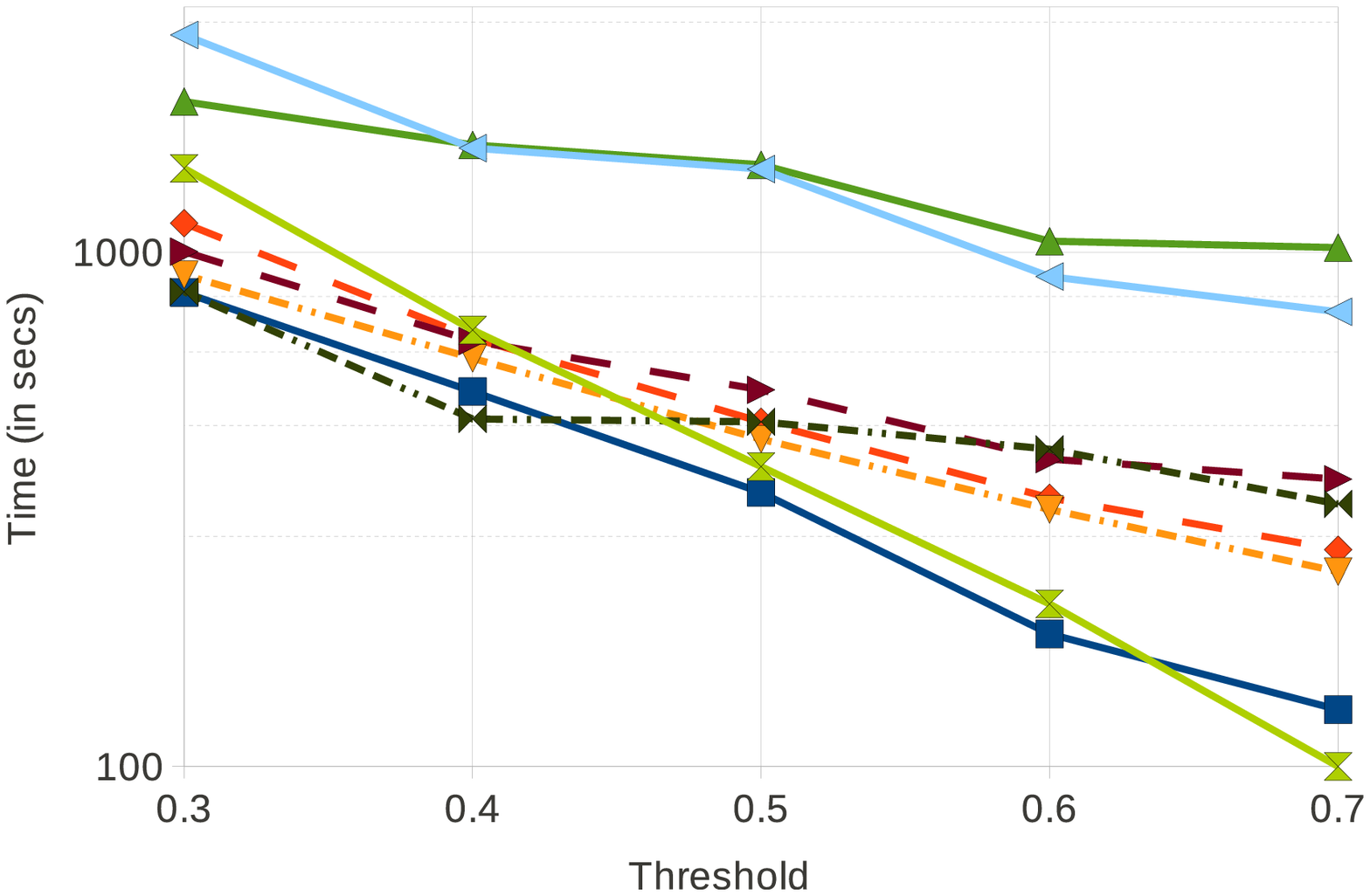}
\label{fig:orkut_jac}
}
\subfigure[Twitter (Binary, Jaccard)]
{
\includegraphics[width=160pt,height=135pt]{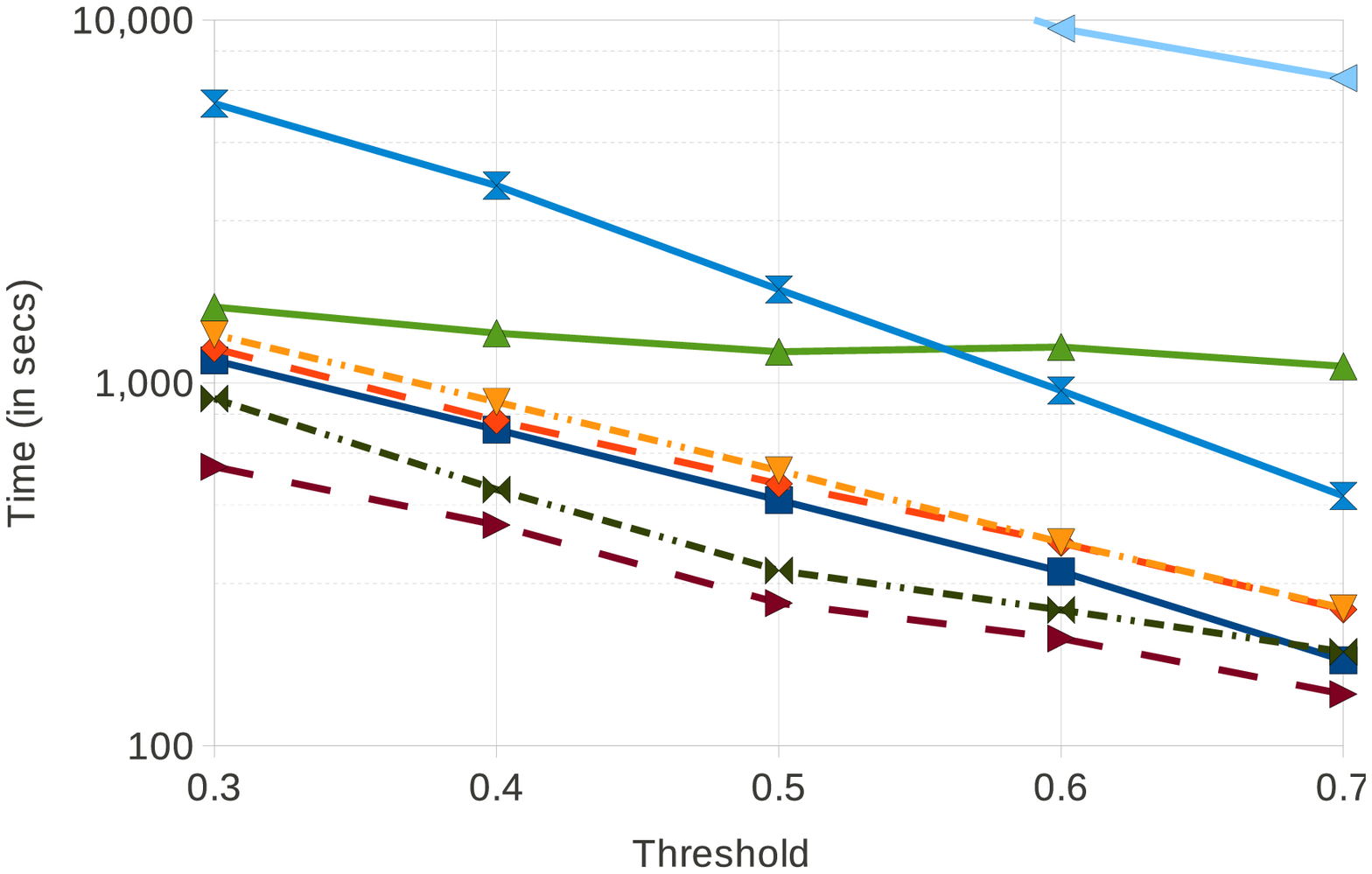}
\label{fig:twitter_jac}
}
\end{center}
\vspace{-0.28in}
\begin{center}
\subfigure[WikiWords500K (Binary, Cosine)]
{
\includegraphics[width=160pt,height=135pt]{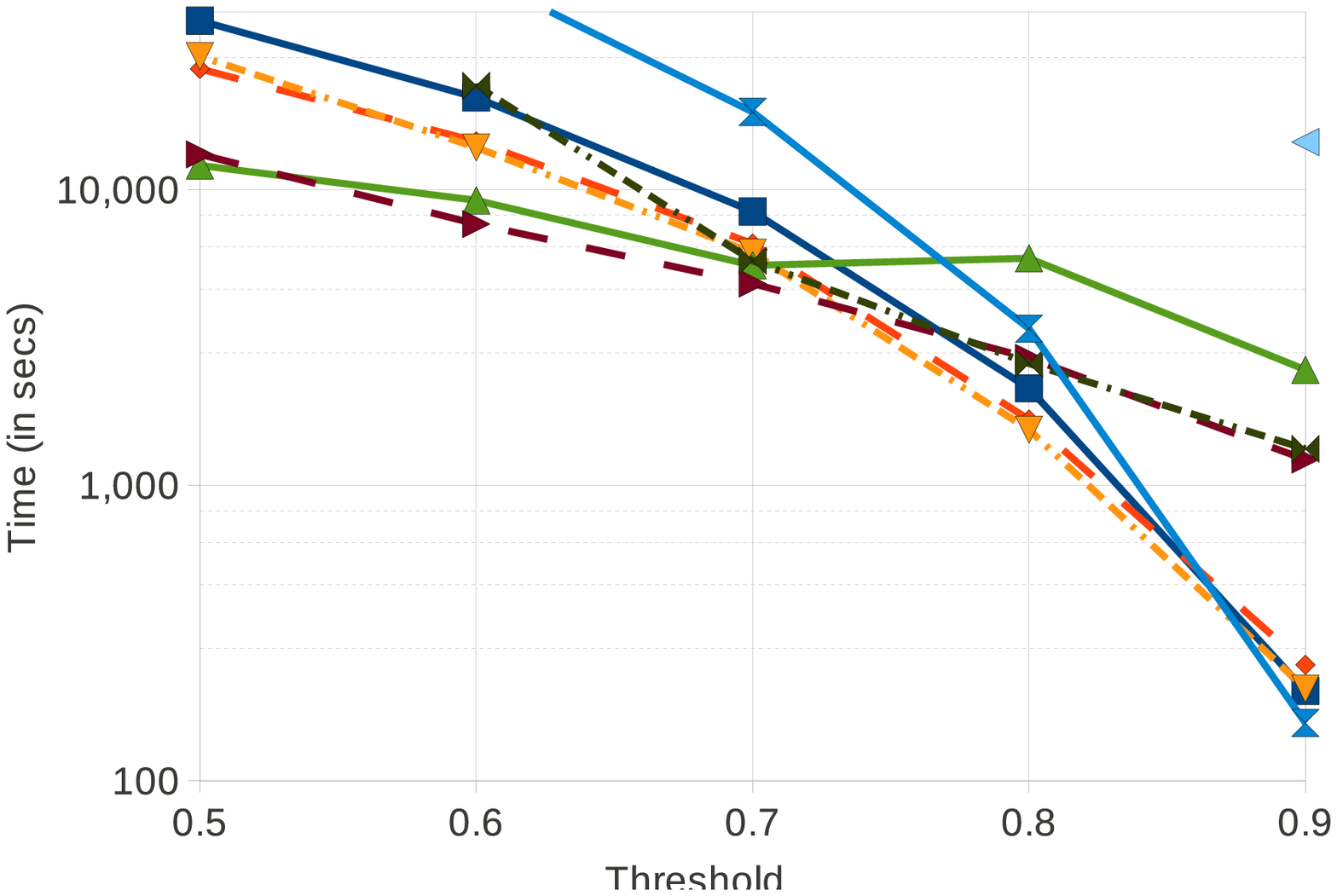}
\label{fig:wiki500k_binCos}
}
\subfigure[Orkut (Binary, Cosine)]
{
\includegraphics[width=160pt,height=135pt]{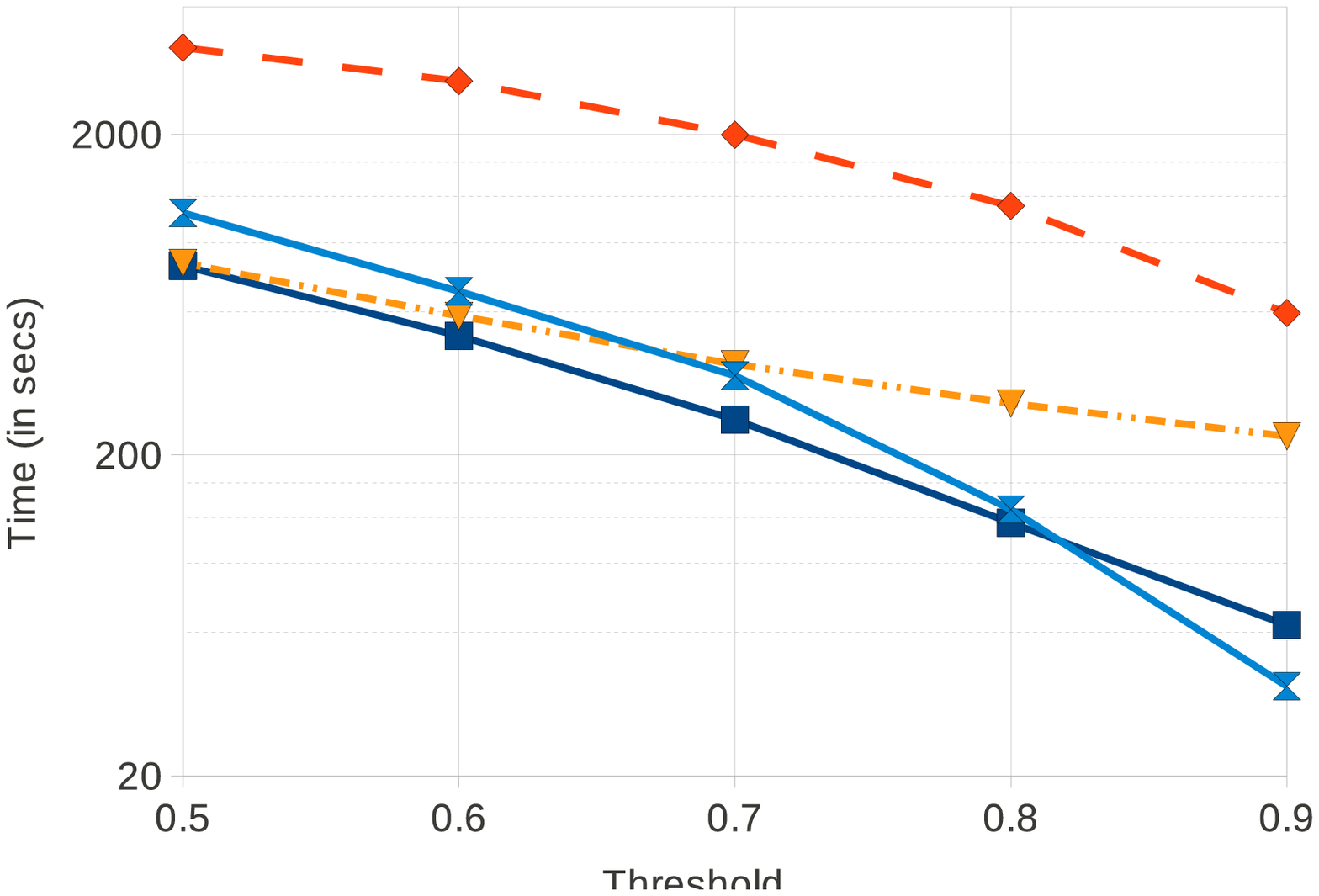}
\label{fig:orkut_binCos}
}
\subfigure[Twitter (Binary, Cosine)]
{
\includegraphics[width=160pt,height=135pt]{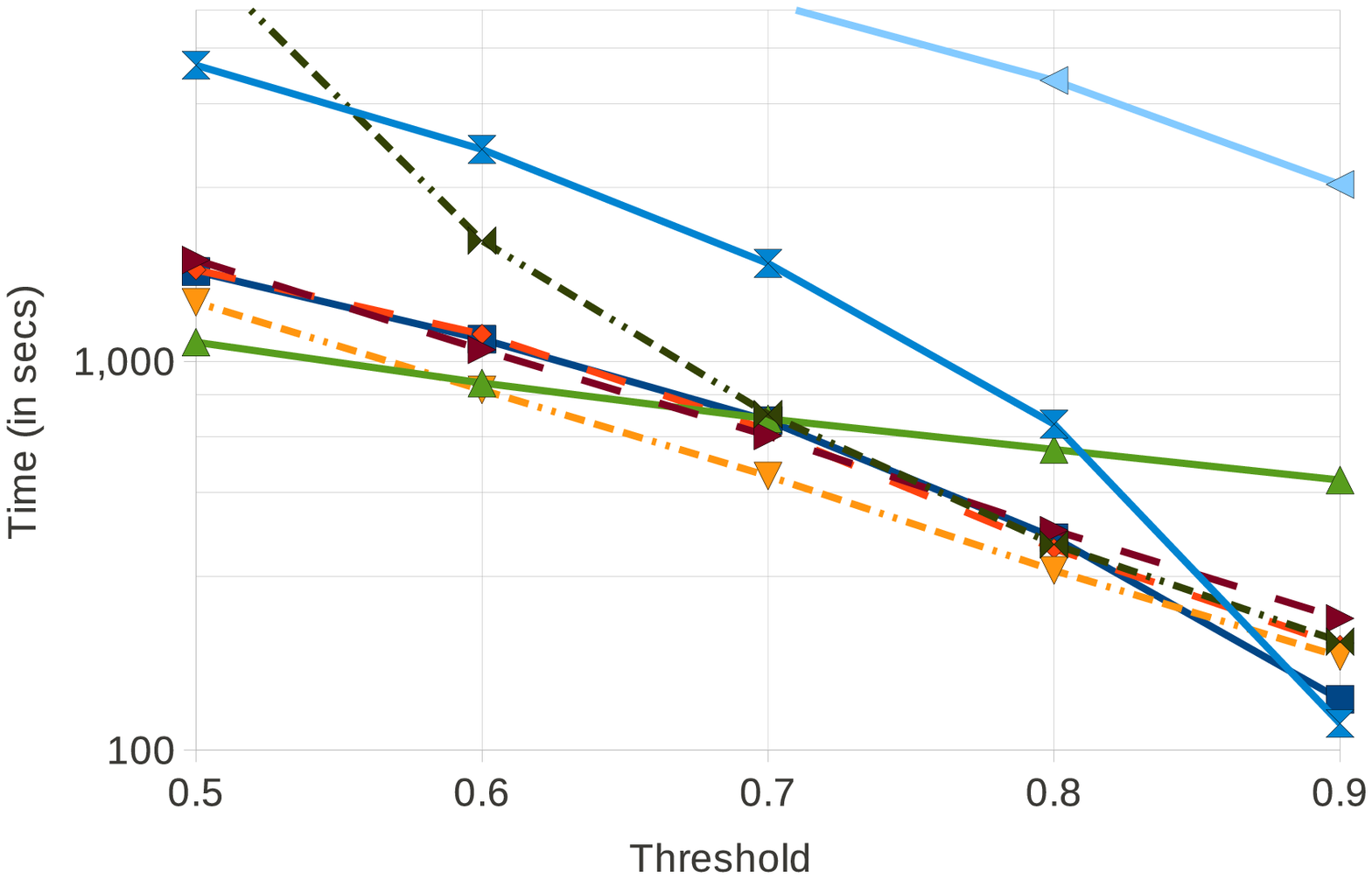}
\label{fig:twitter_binCos}
}
\end{center}
\vspace{-0.28in}
\begin{small}
\caption{Timing comparisons between different algorithms.  
Missing
lines/points are due to the respective algorithm not finishing
within the allotted time (50 hours). 
}
\label{fig:timing}
\end{small}
\end{figure*}

Figure~\ref{fig:timing} shows a comparison of timing results for
all algorithms across a variety of datasets and thresholds.
Table~\ref{tab:bestSpeedups} compares the fastest BayesLSH
variant with all the baselines. 
The quality of the output of BayesLSH can be seen in
Table~\ref{tab:recall} where we show the recall rates for
AP+BayesLSH and AP+BayesLSH-Lite, and in Table~\ref{tab:accuracy} where we
compare the accuracies of LSH and LSH+BayesLSH. The recall and
accuracies of the other BayesLSH variants follow similar
trends and are omitted. The main trends
from the results are distilled and discussed below:

\vspace{0.05in}
\noindent {\bf 1.}
BayesLSH and BayesLSH-Lite improve the running time of both
AllPairs and LSH in almost all the cases, with speedups
usually in the range {\bf 2x-20x}. It can be seen from
Table~\ref{tab:bestSpeedups} that a BayesLSH variant is the
fastest algorithm (in terms of total time across all thresholds) 
for the majority of datasets and similarities, with the exception
of Orkut for Jaccard and binary cosine.
Furthermore, the quality of
BayesLSH output is high; the recall rates are usually above 97\% (see
Table~\ref{tab:recall}), and similarity estimates are accurate,
with usually no more than 5\% output pairs with error above 0.05
(see Table~\ref{tab:accuracy}). 

\begin{figure*}
\begin{center}
\subfigure[WikiWords100K, t=0.7, Cosine]
{
\includegraphics[width=160pt,height=140pt]{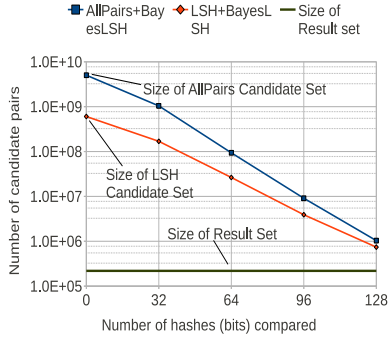}
\label{fig:pruneWiki100K}
}
\hspace{-0.1in}
\subfigure[WikiLinks, t=0.7, Cosine]
{
\includegraphics[width=160pt,height=140pt]{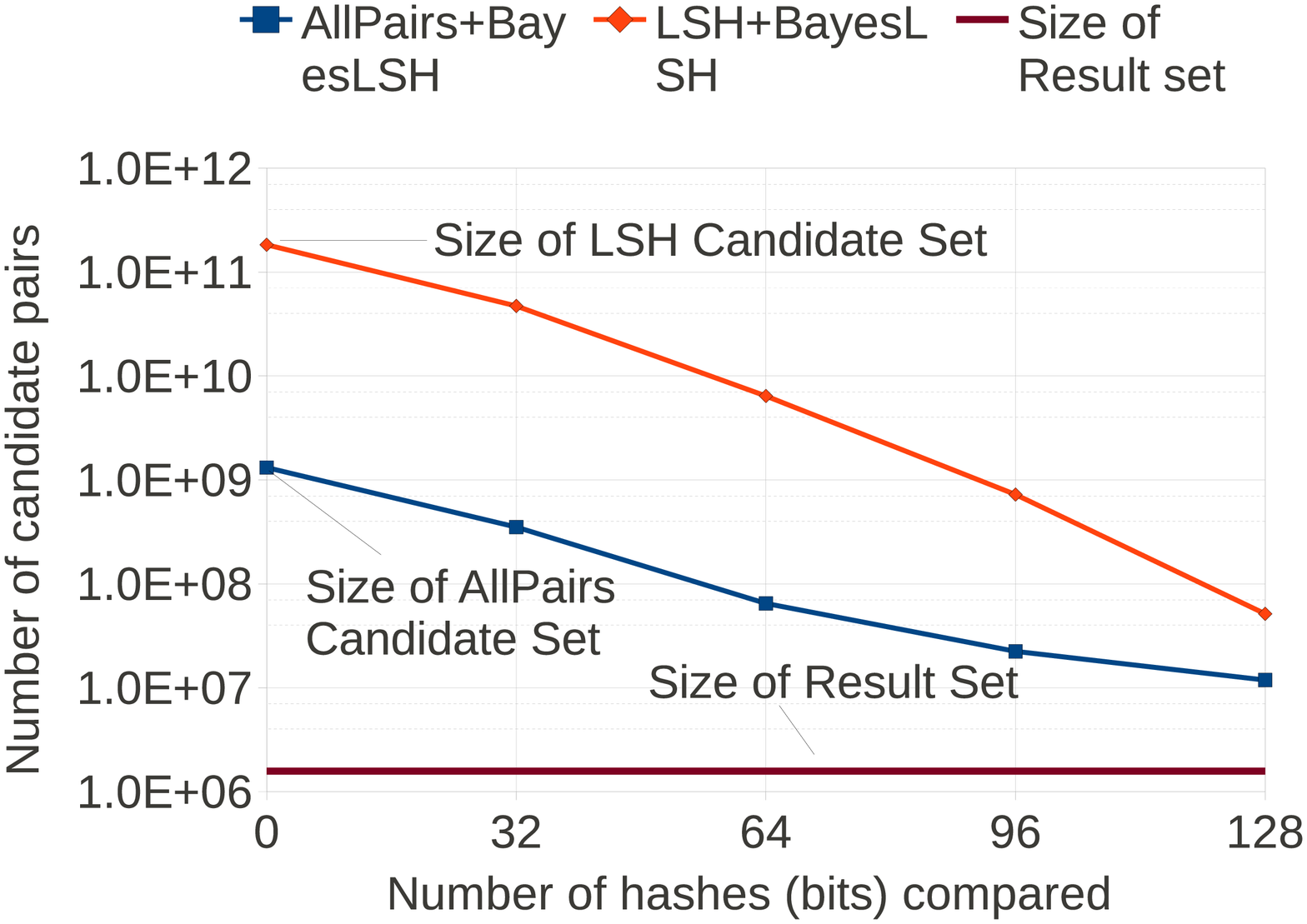}
\label{fig:pruneWikiLinks}
}
\hspace{-0.1in}
\subfigure[WikiWords100K, t=0.7, Binary Cosine]
{
\includegraphics[width=160pt,height=140pt]{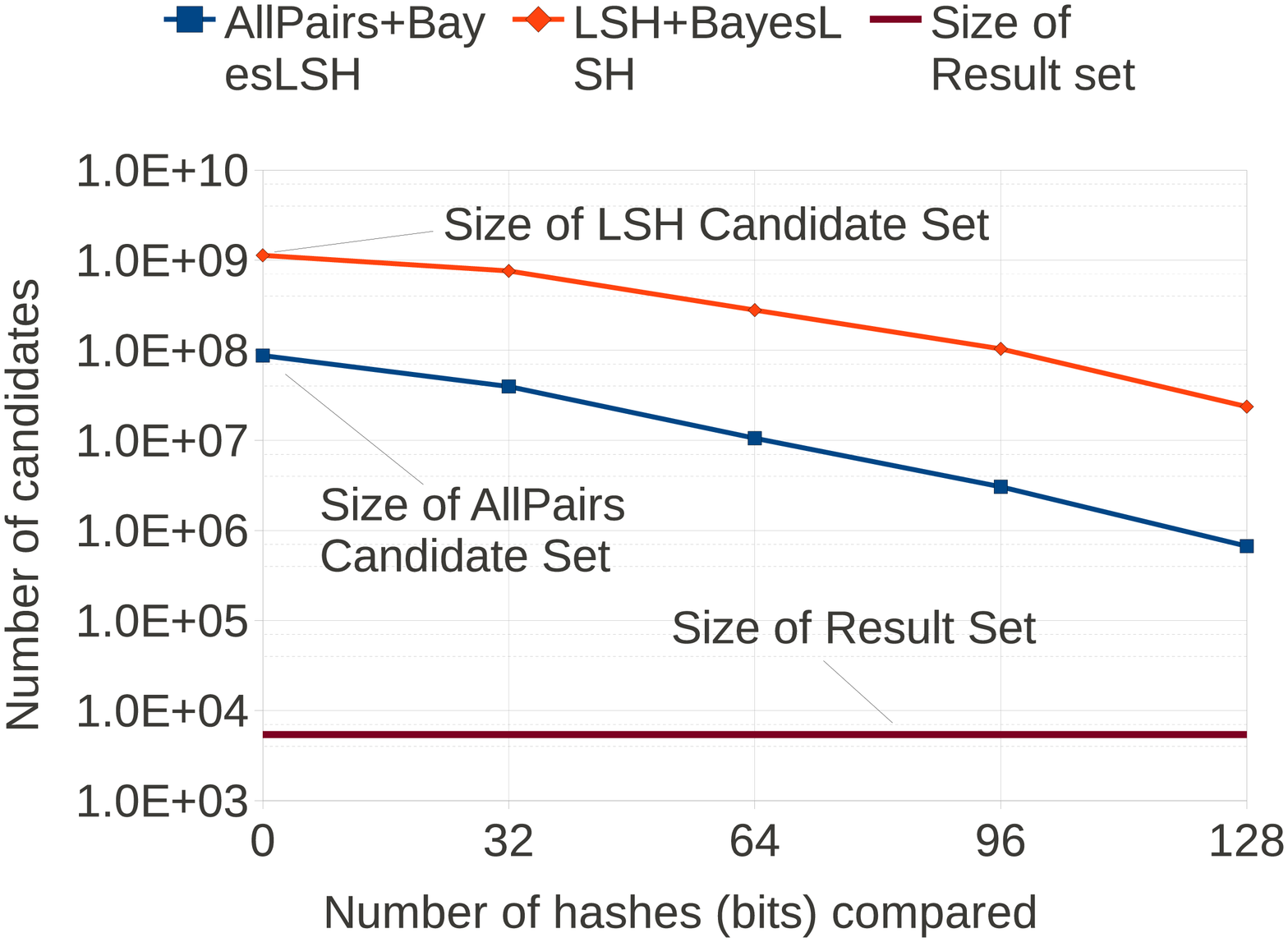}
\label{fig:pruneWiki100K_binCos}
}
\end{center}
\vspace{-0.3in}
\caption{BayesLSH can prune the vast majority of false positive
candidate pairs by examining only a small number of hashes,
resulting in major gains in the running time.}
\label{fig:prune}
\end{figure*}

\vspace{0.05in}
\noindent {\bf 2.}
BayesLSH is fast primarily
by being able to prune away the vast majority of false positives
after comparing only a few hashes. This is illustrated in
Figure~\ref{fig:prune}. For WikiWords100K at a threshold of 0.7,
(see Figure~\ref{fig:pruneWiki100K})
AllPairs supplies BayesLSH with nearly 5e09 candidates, while
the result set only has 2.2e05. BayesLSH is able to prune away
{\bf 4.0e+09} (80\%) of the input candidate pairs 
after examining only 32 hashes
- in this case, each hash is a bit, so BayesLSH compared only 4
bytes worth of hashes between each pair. By the time BayesLSH has
compared 128 hashes (16 bytes) there are only 1.0e06
candidates remaining. Similarly LSH supplies BayesLSH with
6.0e08 candidates - better than AllPairs, but nonetheless
orders of magnitude larger than the final result set - and
after comparing 128 hashes (16 bytes), BayesLSH is able to prune 
that down to only 7.4e05, only about 3.5x larger than the
result set. On the WikiLinks dataset (see
Figure~\ref{fig:pruneWikiLinks}), we see a similar trend with the
roles of AllPairs and LSH reversed - this time it is AllPairs
instead which supplies BayesLSH with fewer candidates. 
After examining only 128 hashes, BayesLSH
is able to reduce the number of candidates from 1.3e09 down to
1.2e07 for AllPairs, and from 1.8e11 down to 5.1e07 for LSH.
Figure~\ref{fig:pruneWiki100K_binCos} shows a similar trend, this
time on the binary version of WikiWords100K.

\vspace{0.05in}
\noindent {\bf 3.} We note that
BayesLSH and BayesLSH-Lite often (but not always) have comparable
speeds, since most of the speed benefit is coming from the
ability of BayesLSH to prune, which is an aspect that is common
to both algorithms. The difference between the two is mainly in
terms of the hashing overhead. BayesLSH needs to obtain many more
hashes of each object in order for similarity estimation; 
this cost is amortized at lower
thresholds, where the number of similarity calculations needed to
perform is much greater. BayesLSH-Lite is faster at higher
thresholds or when exact similarity calculations are cheaper,
such as datasets with low average vector length.

\vspace{0.05in}
\noindent {\bf 4.}
AllPairs and LSH have complementary strengths and weaknesses. On
the datasets RCV1, WikiWords100K, WikiWords500K and Twitter (see
Figures~\ref{fig:rcv1}-\ref{fig:wiki500k},\ref{fig:twitter}), LSH
is clearly the faster algorithm than AllPairs (in the case of
WikiWords500K, AllPairs did not finish execution even for the
highest threshold of 0.9). On the other hand, AllPairs is the 
much faster algorithm on WikiLinks and Orkut (see
Figures~\ref{fig:wikiLinks}-\ref{fig:orkut}), with LSH timing out
in most cases. Looking at the characteristics of the datasets,
one can discern a pattern: AllPairs is faster on datasets with
smaller average length and greater variance in the vector
lengths, as is the case with the graph datasets WikiLinks and
Orkut. The variance in the vector lengths allows AllPairs to
upper-bound the similarity better and thus prune away
more false positives, and in addition the exact similarity
computations that AllPairs does are faster when the average
vector length is smaller. 
However, BayesLSH and BayesLSH-Lite enable speedups on \emph{both}
AllPairs and LSH, not only when each algorithm is slow, but even
when each algorithm is \emph{already fast}.

\vspace{0.05in}
\noindent {\bf 5.}
The accuracy of BayesLSH's similarity estimates is much more
consistent as compared to the standard LSH approximation, as can
be seen from Table~\ref{tab:accuracy}. 
LSH generally produces too many errors
when the threshold is low and too few errors when the threshold
is high. This is mainly because LSH uses the same number of
hashes (set to 2048) for estimating all similarities, low and
high. This problem would persist even if the number of hashes was
set to some other value, as explained in
Section~\ref{subsub:manyhashes}. BayesLSH, on the other hand,
maintains similar
accuracies at both low and high thresholds, \emph{without requiring any
tuning at all on the number of hashes to be compared}, and only
based on the user's specification of the desired accuracy using
$\delta, \gamma$ parameters. 

\vspace{0.05in}
\noindent {\bf 6.}
LSH Approx is often much faster than LSH with exact similarity
calculations, especially for datasets with higher average vector
lengths, where the speedup is often 3x or more - on Twitter, the
speedup is as much as 10x (see Figure~\ref{fig:twitter}). 

\vspace{0.05in}
\noindent {\bf 7.}
BayesLSH does not enable speedups that are as significant for
AllPairs in the case of binary vectors. We found that this was
because AllPairs was already doing a very good job at generating
a small candidate set, thus not leaving much room for
improvement. In contrast, LSH was still generating a large
candidate set, leaving room for LSH+BayesLSH to enable speedups.
Interestingly, even though LSH generates about 10 times more candidates
than AllPairs, the LSH variants of BayesLSH are about 50-100\%
\emph{faster} than AllPairs and its BayesLSH versions, on
WikiWords500K and Twitter (see
Figures~\ref{fig:wiki500K_jac},\ref{fig:twitter_jac}). This is
because LSH is a faster indexing and candidate generation
strategy, especially when the average vector length is large.

\vspace{0.05in}
\noindent {\bf 8.}
PPJoin+ is often the fastest algorithm at the highest thresholds
(see Figures~\ref{fig:wiki500K_jac}-\ref{fig:twitter_binCos}),
but its performance degrades very rapidly with lower thresholds.
A possible explanation is that the pruning heuristics used in
PPJoin+ are effective only at higher thresholds.

\subsection{Effect of varying parameters of BayesLSH}
\begin{figure}
\begin{center}
\includegraphics[width=200pt,height=180pt]{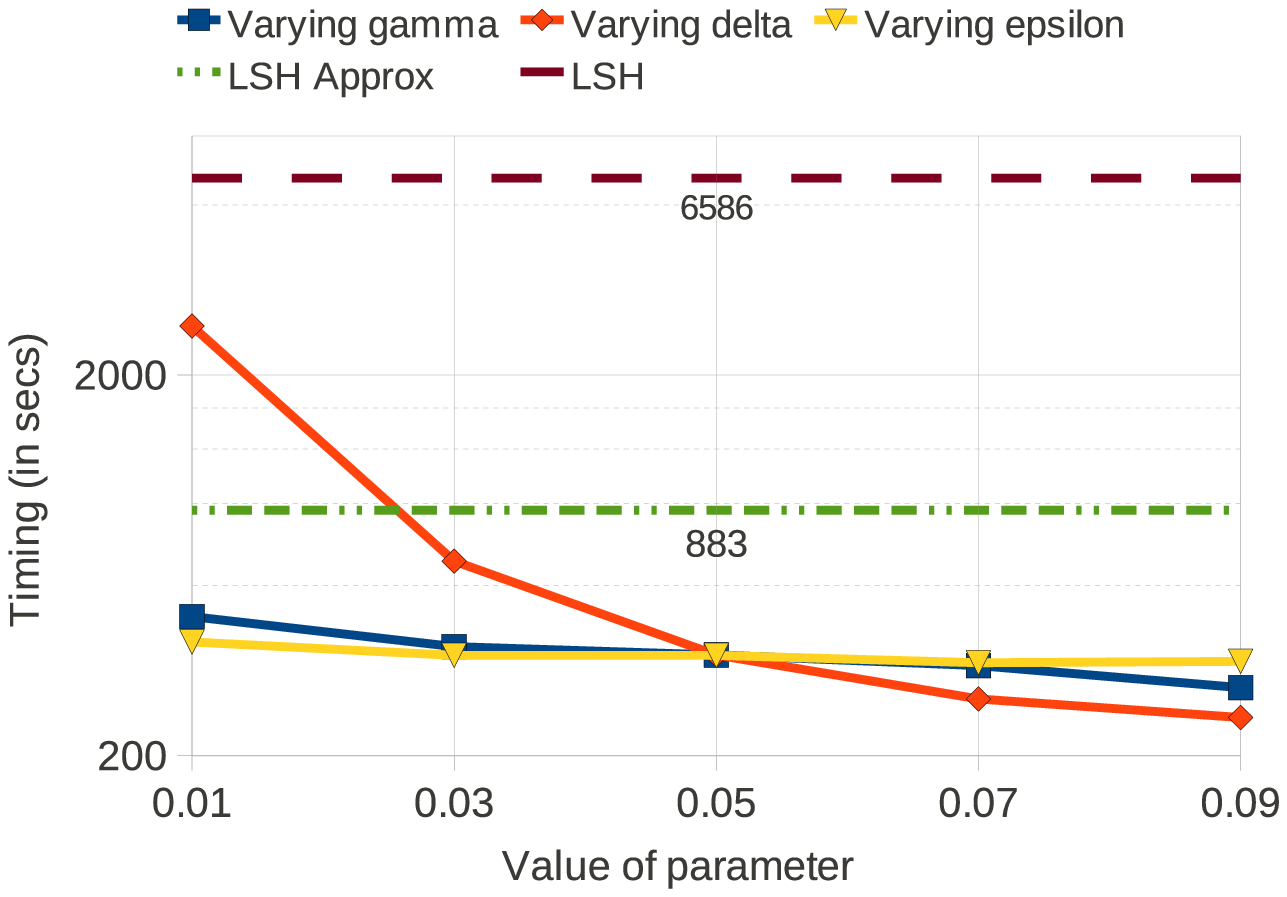}
\end{center}
\vspace{-0.4in}
\caption{Effect of varying $\gamma, \delta, \epsilon$
separately on the running time of LSH+BayesLSH. The dataset is WikiWords100K, with the
threshold fixed at t=0.7 for cosine similarity. 
For comparison, the times for LSH Approx and LSH are also shown.}
\label{fig:varyPars}
\end{figure}

We next examine the effect of varying the parameters of BayesLSH
- namely the accuracy parameters $\gamma, \delta$ and the recall
parameter $\epsilon$. We vary each parameter from 0.01 to 0.09 in
increments of 0.02, while fixing the other two parameters to
0.05, and fix the dataset to WikiWords100K and threshold to 0.7
(cosine similarity). 
The effect of varying each of these parameters on the
execution time is plotted in Figure~\ref{fig:varyPars}. 
Varying the 
recall parameter $\epsilon$ and the accuracy
parameter $\gamma$ have barely any effect on the running time -
however setting $\delta$ to lower values does increase the
running time significantly. Why does lowering $\delta$ penalize
the running time much more than lowering $\gamma$? This is
because lowering $\delta$ increases the number of hashes that
have to be compared for \emph{all} result pairs, while lowering
$\gamma$ increases the number of hashes that have to be compared
only for those result pairs that have uncertain similarity
estimates. It is interesting to note that even though
$\delta=0.01$ requires 2691 secs, it achieves a very low mean
error of 0.001, while being much faster than
LSH exact, which requires 6586 secs. Approximate LSH requires 883
secs but is much more error-prone, with a mean error of 0.014.
With $\gamma=0.01$, BayesLSH achieves a mean error of 0.013,
while still being around 2x faster than approximate LSH.

In Table~\ref{tab:vary}, we show the result of varying these
parameters on the output quality. When varying a 
parameter, we show the change in output quality only for the
relevant quality metric - e.g. for changing $\gamma$ we only
show how the fraction of errors $>$ 0.05 changes, since we find
that recall is largely unaffected by changes in $\gamma$ and
$\delta$ (which is as it should be). 
Looking at the column corresponding to varying
$\gamma$, we find that the fraction of errors $>$ 0.05 increases as
we expect it to when we increase $\gamma$, without ever exceeding
$\gamma$ itself. When varying $\delta$, we can see that the mean
error reduces as expected for lower values of $\delta$. Finally,
when varying the recall parameter $\epsilon$, we find that the
recall reduces with higher values of $\epsilon$ as expected, with
the false negative rate always less than $\epsilon$ itself.

\begin{table}
\begin{center}
\begin{scriptsize}
\hspace{-0.3in}
\begin{tabular}{|p{0.6in}|p{0.5in}|c|c|p{0.3in}|c|}
\hline
\multirow{2}*{Dataset} & \multirow{2}{0.5in}{Fastest BayesLSH variant} 
& \multicolumn{4}{c|}{Speedup
w.r.t baselines} \\
\cline{3-6}
& & AP & LSH & LSH Approx & PPJoin \\
\hline
\multicolumn{6}{|c|}{{\bf Tf-Idf, Cosine}} \\
\hline
RCV1 & LSH + BayesLSH & 7.1x & 4.8x & 2.4x & - \\
\hline
WikiWords-100K & LSH + BayesLSH & 31.4x & 15.1x & 2.0x & - \\
\hline
WikiWords-500K & LSH + BayesLSH & $\geq$ 42.1x & $\geq$ 13.3x & 2.8x & - \\
\hline
WikiLinks & AP + BayesLSH-Lite & 1.8x & $\geq$ 248.2x & $\geq$
246.3x & - \\
\hline
Orkut & AP + BayesLSH-Lite & 1.2x & $\geq$ 114.9x & $\geq$ 155.6x
& - \\
\hline
Twitter & LSH + BayesLSH & 26.7x & 33.4x & 3.0x & - \\
\hline
\multicolumn{6}{|c|}{{\bf Binary, Jaccard}} \\
\hline
WikiWords-500K & LSH + BayesLSH & 2.0x & $\geq$ 16.8x & 3.7x &
5.2x\\
\hline
Orkut & AP + BayesLSH-Lite & 0.8x & 2.9x & 2.8x & 1.1x \\
\hline
Twitter & LSH + BayesLSH & 1.8x & 48.4x & 4.2x & 8.0x \\
\hline
\multicolumn{6}{|c|}{{\bf Binary, Cosine}} \\
\hline
WikiWords-500K & LSH + BayesLSH & 2.3x & $\geq$ 10.2x & 1.2x &
5.6x \\
\hline
Orkut & AP + BayesLSH-Lite & 0.8x & $\geq$ 201x & $\geq$ 201x &
1.0x \\
\hline
Twitter & AP + BayesLSH-Lite & 1.2x & 27.4x & 1.2x & 3.7x \\
\hline
\end{tabular}
\end{scriptsize}
\caption{Fastest BayesLSH variant for each dataset (based on total time
across all thresholds), and speedups over each baseline. BayesLSH
variants are fastest in all cases except for binary versions of
Orkut, where it is only slightly sub-optimal. The range of
thresholds for Cosine was 0.5 to 0.9, and for Jaccard was 0.3 to
0.7. PPJoin is only applicable to binary datasets. 
In some cases, only lower-bound on speedup is available as
the baselines timed out (indicated with $\geq$).}
\label{tab:bestSpeedups}
\end{center}
\end{table}

\begin{table}
\begin{small}
\vspace{-0.4in}
\begin{tabular}{|c|c|c|c|c|c|}
\hline
{\bf Dataset} & t=0.5 & t=0.6 & t=0.7 & t=0.8 & t=0.9 \\
\hline
\multicolumn{6}{|c|}{{\bf AllPairs+BayesLSH}} \\
\hline
RCV1 & 97.97 & 98.18 & 98.47 & 99.08 & 99.36 \\
\hline
WikiWords100K & 98.52 &	98.84 &	99.2 &	98.58 &	96.69 \\
\hline
WikiWords500K & 97.54  & 97.82 & 98.21 & 98.16 & 96.66 \\
\hline
WikiLinks & 97.45 & 98.04 & 98.46 & 98.68 & 99.18 \\
\hline
Orkut & 97.1 & 97.8 & 98.86 & 99.84 & 99.99 \\
\hline
Twitter & 97.7 & 96 & 96.88 & 97.33 & 98.77 \\
\hline
\multicolumn{6}{|c|}{{\bf AllPairs+BayesLSH-Lite}} \\
\hline
RCV1 & 98.73 & 98.82 & 98.89 & 99.26 & 99.55 \\
\hline
WikiWords100K & 98.88 & 99.31 & 99.62 & 99.69 & 99.5 \\
\hline
WikiWords500K & 98.79 & 98.72  & 98.98 & 98.74 & 98.83 \\
\hline
WikiLinks & 98.53 & 98.91 & 99.16 & 99.18 & 99.45 \\
\hline
Orkut & 98.4 & 98.64 & 99.3 & 99.87 & 99.99 \\
\hline
Twitter & 99.44 & 98.82 & 97.17 & 97.18 & 99.06 \\
\hline

\end{tabular}
\end{small}
\caption{Recalls (out of 100) 
of AllPairs+BayesLSH and AllPairs+BayesLSH-Lite across 
different datasets and different similarity thresholds.}
\label{tab:recall}
\end{table}

\begin{table}
\begin{small}
\begin{tabular}{|c|c|c|c|c|c|}
\hline
 & t=0.5 & t=0.6 & t=0.7 & t=0.8 & t=0.9 \\
\hline
\multicolumn{6}{|c|}{{\bf LSH Approx}} \\
\hline
RCV1 & 7.8 & 4.3 & 2.25 & 0.8 & 0.04 \\
\hline
WikiWords100K &4.7 & 3.6 & 1 & 0.3 & 0.02 \\
\hline
WikiWords500K &8.3 & 5.7 & 2.9 & 0.9 & 0.1 \\
\hline
WikiLinks & - & - & 1.6 & 0.4 & 0.06 \\
\hline
Orkut & - & - & - & - & 0.0072 \\
\hline
Twitter & 4 & 5.1 & 2.6 & 0.4 & 0.02 \\
\hline
\multicolumn{6}{|c|}{{\bf LSH + BayesLSH}} \\
\hline
RCV1 & 3.2 & 2.9 & 3.2 & 2 & 1.4 \\
\hline
WikiWords100K &2.7 & 2.3 & 3.5 & 4.9 & 2.2 \\
\hline
WikiWords500K &3.4 & 3.4 & 3.2 & 2.9 & 2.1\\
\hline
WikiLinks & 2.96 & 2.82 & 2.3 & 2 & 1.6 \\
\hline
Orkut & - & - & 1.5 & 0.6 & 0.09 \\
\hline
Twitter & 2.3 & 4 & 3.1 & 4.8 & 4.3 \\
\hline
\end{tabular}
\end{small}
\caption{Percentage of similarity estimates with errors greater than $0.05$; 
comparison between LSH Approx and LSH + BayesLSH}
\label{tab:accuracy}
\end{table}

\begin{table}
\begin{tabular}{|c|p{0.8in}|p{0.6in}|p{0.6in}|}
\hline
Parameter value & Fraction errors $>$
0.05 for varying $\gamma$ & Mean
error for varying $\delta$ & Recall for varying $\epsilon$ \\
\hline
0.01 & 0.7\% & 0.001 &  98.76\% \\
\hline
0.03 & 2\% & 0.01 & 97.79\% \\
\hline
0.05 & 3\% & 0.017 & 97.33\% \\
\hline
0.07 & 4.2\% & 0.022 & 96.06\% \\
\hline
0.09 & 5.4\% & 0.027 & 95.35\% \\
\hline
\end{tabular}
\caption{The effect of varying the parameters $\gamma,
\delta, \epsilon$ one at a time, while fixing the other two parameters at
0.05. The dataset is WikiWords100K, with the threshold fixed at
t=0.7; the candidate generation algorithm was LSH.}
\vspace{-0.2in}
\label{tab:vary}
\end{table}

\section{Conclusions and Future Work}
In this article, we have presented BayesLSH 
(and a simple variant
BayesLSH-Lite), a general candidate
verification and similarity estimation algorithm for approximate
similarity search, which combines Bayesian inference with LSH in
a principled manner and has a number of advantages compared to
standard similarity estimation using LSH. 
BayesLSH enables significant speedups for
two state-of-the-art candidate generation algorithms, AllPairs
and LSH, across a wide variety of datasets, and furthermore the
quality of BayesLSH is easy to tune. As can be seen from
Table~\ref{tab:bestSpeedups}, a BayesLSH variant is typically the fastest
algorithm on a variety of datasets and similarity measures.

BayesLSH takes a largely orthogonal direction to a lot of recent
research in LSH, which concentrates on more effective indexing
strategies, ultimately with the goal of candidate generation,
such as Multi-probe LSH~\cite{lv07} and LSB-trees~\cite{tao09}.
Furthermore, a lot of research on LSH is concentrated on
nearest-neighbor retrieval for distance measures, rather than all
pairs similarity search with a similarity threshold $t$. 

There are two promising avenues for future research with
BayesLSH. First is to extend BayesLSH for similarity search with
learned (kernelized) metrics, since such similarity measures are
often superior for complex domains~\cite{jain08}. Secondly, we
believe that a BayesLSH-Lite analogue can be developed 
for candidate pruning in the
case of nearest neighbor retrieval for Euclidean 
distances
(although the final distance may have to be calculated exactly). 

\vspace{0.05in}
\noindent
{\bf Acknowledgments:} We thank Luis Rademacher and anonymous
reviewers for helpful
comments, and Roberto Bayardo for clarifications on the AllPairs
implementation. This work is supported in
part by the following NSF grants: IIS-1141828 and IIS-0917070.

\begin{scriptsize}
\bibliographystyle{abbrv}
\bibliography{paper}
\end{scriptsize}

\pagebreak
\appendix

\section{The Influence of Prior vs. Data}

\begin{figure}
\begin{center}
\hspace{-0.3in}
\subfigure[Prior distributions]
{
\includegraphics[width=120pt,height=100pt]{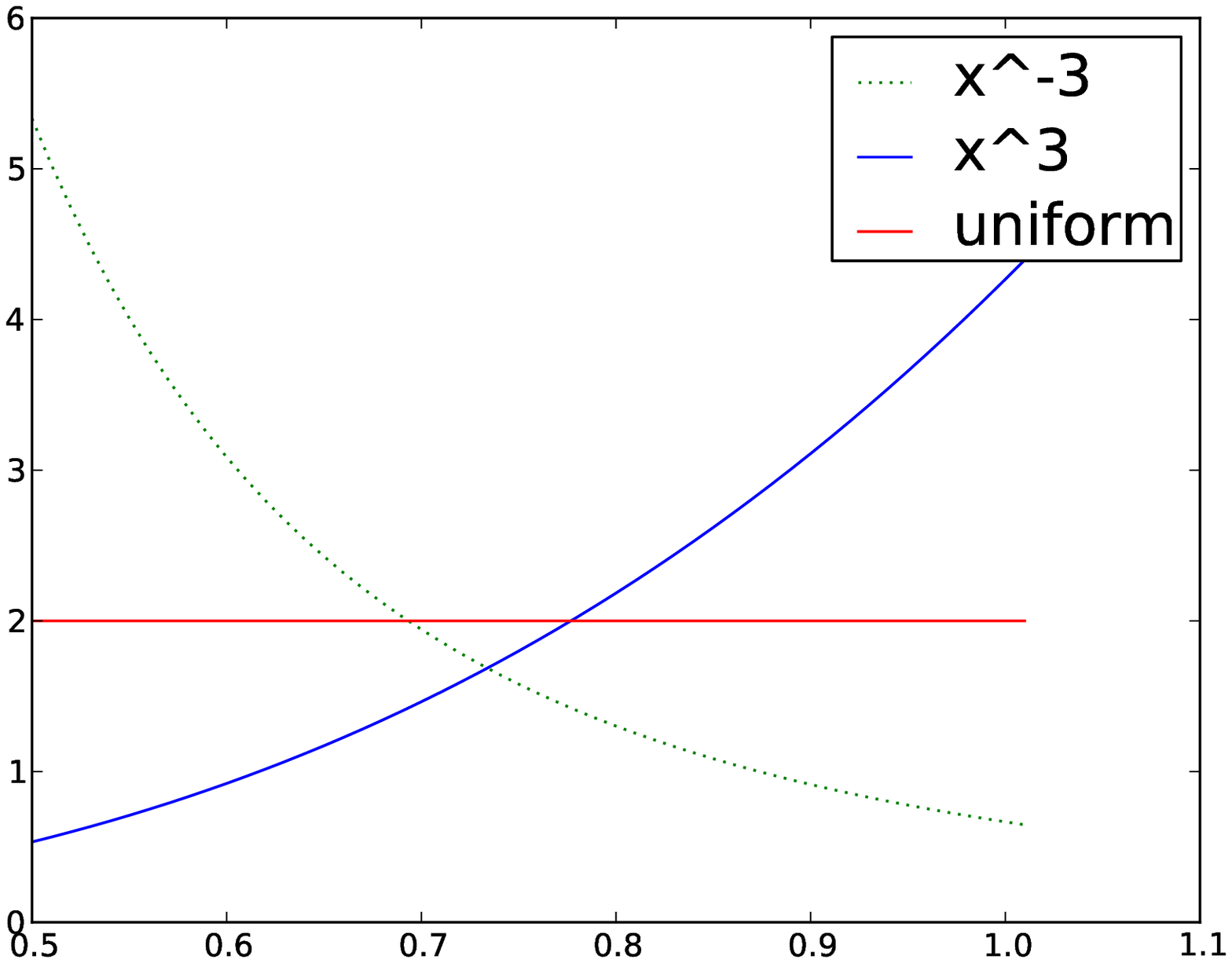}
\label{fig:m0n0}
}
\hspace{-0.1in}
\subfigure[Posterior after examining 32 hashes, with 24 agreements]
{
\includegraphics[width=120pt,height=100pt]{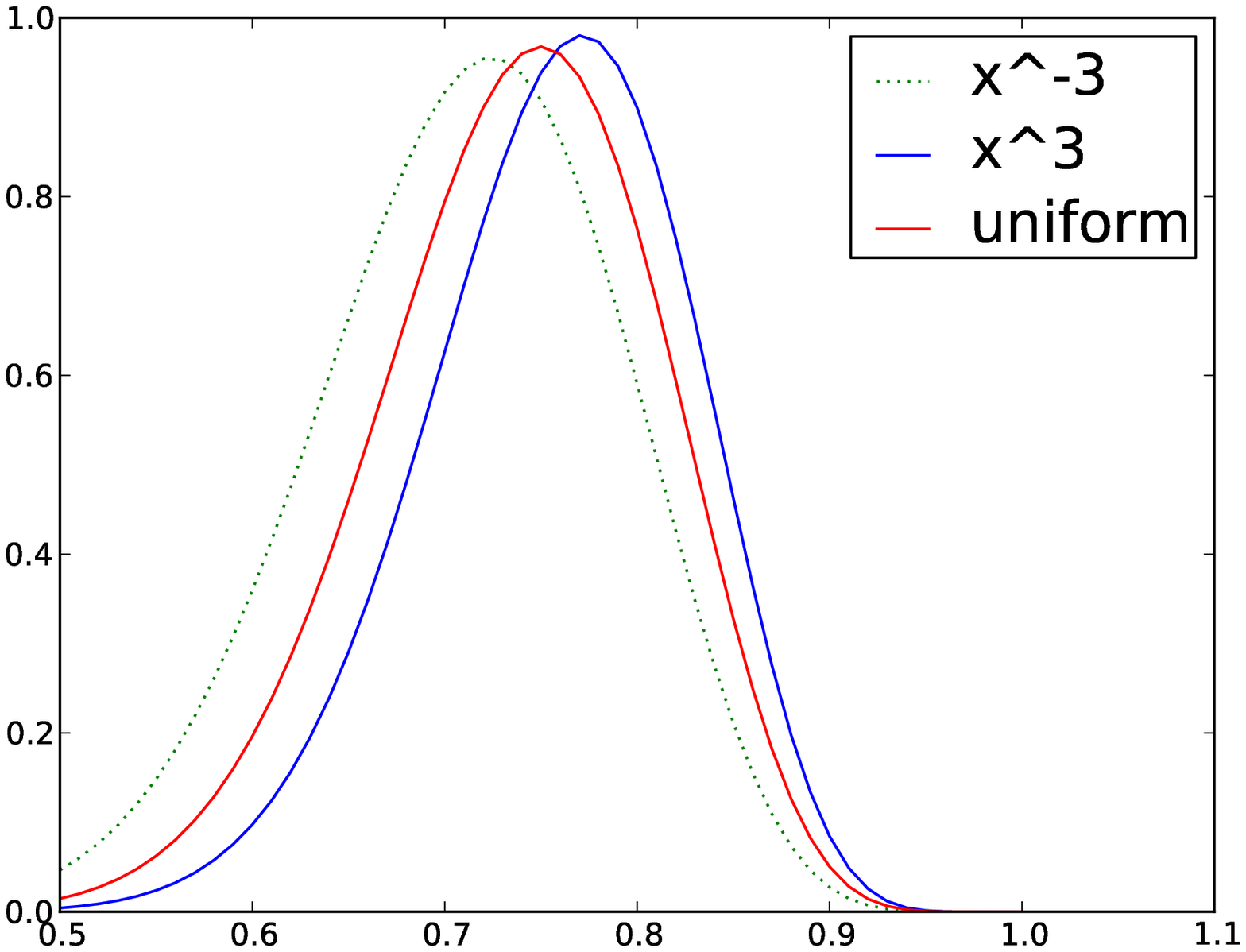}
\label{fig:m24n32}
}
\end{center}
\end{figure}
\begin{figure}
\begin{center}
\vspace{-0.2in}
\hspace{-0.35in}
\subfigure[Posterior after examining 64 hashes, with 48 agreements]
{
\includegraphics[width=120pt,height=100pt]{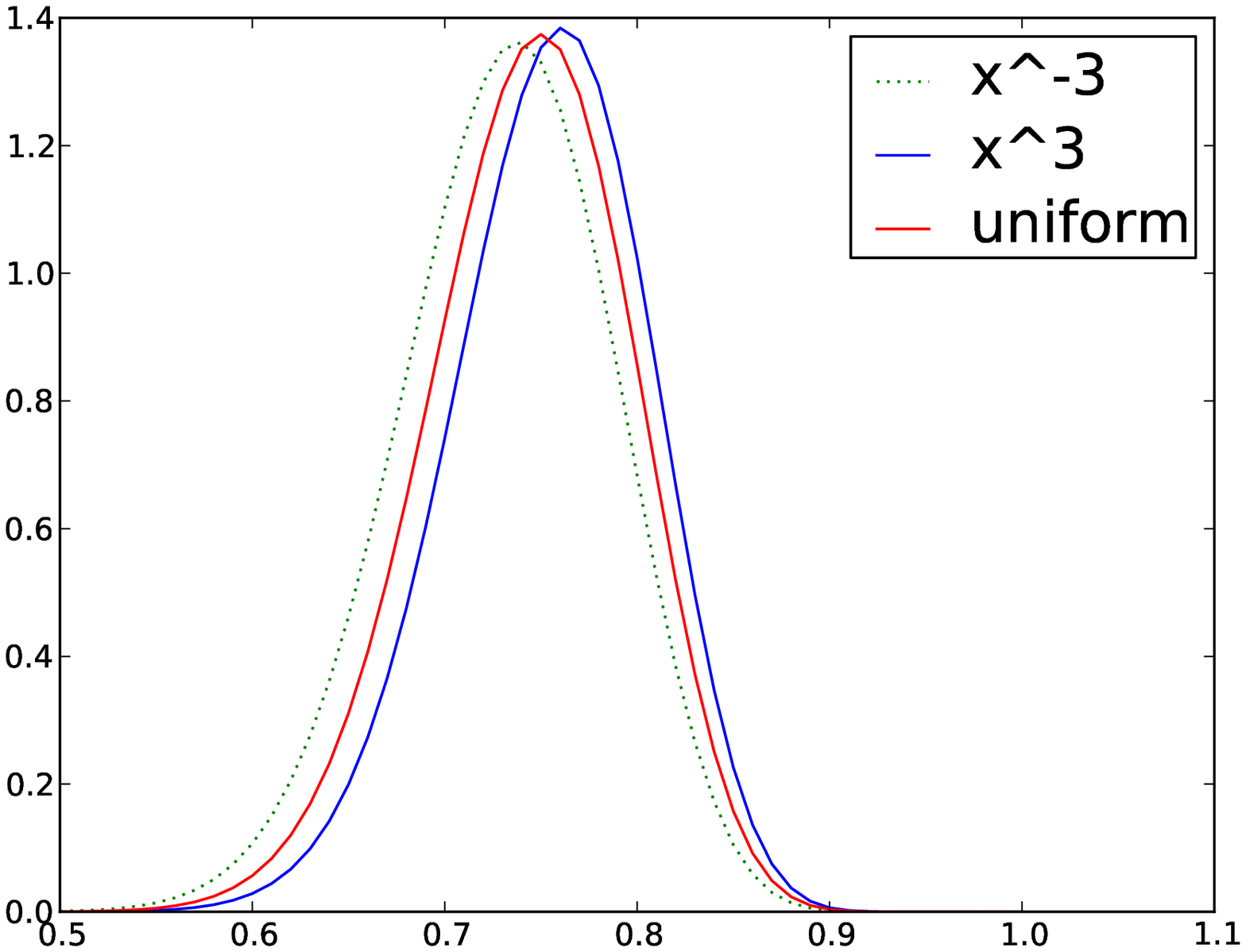}
\label{fig:m48n64}
}
\hspace{-0.05in}
\subfigure[Posterior after examining 128 hashes, with 96 agreements]
{
\includegraphics[width=120pt,height=100pt]{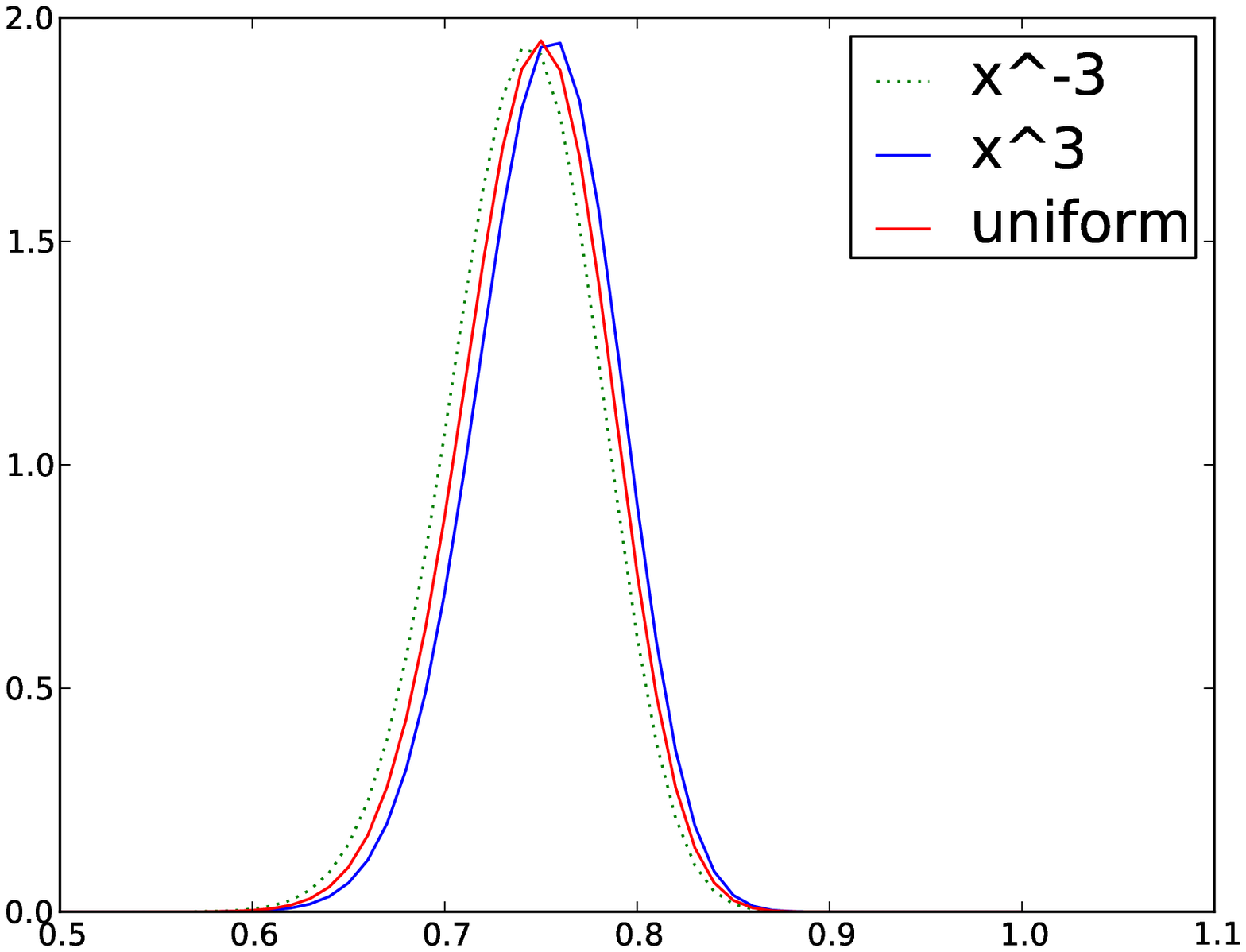}
\label{fig:m96n128}
}
\end{center}
\label{fig:priors}
\vspace{-0.2in}
\caption{Even very different prior distributions converge to very
similar posteriors after examining a small number of outcomes
(hashes)}
\end{figure}

In this section, we show how the observed outcomes (i.e. hashes)
are much more influential in determining the posterior
distribution than the prior itself. Even if we start with very
different prior distributions, the posterior distributions
typically become very similar after observing a surprisingly
small number of outcomes.

Consider the similarity measure we worked with in the case of
cosine similarity, $r(x,y) = 1 - \frac{\theta(x,y)}{\pi}$, which
ranges from [0.5,1] - note that $r(x,y)=0.5$ corresponds to an
actual cosine similarity of 0 between $x,y$. 
Consider three very different prior
distributions for this similarity measure, as follows 
(the normalization constants have been omitted):
\begin{itemize}
\item
Negatively sloped power law prior: $p(s) \propto x^{-3}$
\item
Uniform prior: $p(s) \propto 1$
\item
Positively sloped power law prior: $p(s) \propto x^{3}$
\end{itemize}

In Figure~\ref{fig:priors}, we show the posteriors for each of
these three priors after observing a hypothetical series of
outcomes for a pair of points $x,y$ with cosine similarity 0.70,
corresponding to $r(x,y)=0.75$.  
Although to start off with, the three priors are very
different (see Figure~\ref{fig:m0n0}), the posteriors are already
quite close after observing only 32 hashes and 24 agreements
(Figure~\ref{fig:m24n32}), and
the posteriors get closer quickly with increasing number of
hashes(Figures~\ref{fig:m48n64} and \ref{fig:m96n128}). 

In general, the likelihood term - which, after observing $n$
hashes with $m$ agreements, is $s^m{(1-s)}^{(n-m)}$ - is much
more sharply concentrated than 
a justifiable prior, very quickly as we increase $n$. In other
words, a prior would itself have to be very sharply concentrated
for it to match the influence of the likelihood -
using sharply concentrated priors however brings the danger of
not letting the data speak for themselves. 

\end{document}